\documentclass[preprint,prb,amsmath,amssymb,amsfonts,superscriptaddress,floatfix,showpacs,aps]{revtex4-1}
\usepackage{graphicx}% Include figure files
\usepackage{dcolumn}% Align table columns on decimal point
\usepackage{keyval}
\usepackage{bm}% bold math
\usepackage{subfigure}

\usepackage[usenames,dvipsnames]{color}
\usepackage[normalem]{ulem}

\newcommand{\comment}[1]{\textcolor{red}{#1}}
\renewcommand{\comment}[1]{\relax}

\newcommand{\todelete}[1]{\textcolor{green}{\sout{#1}}}
\renewcommand{\todelete}[1]{\relax}

\newcommand{\newtext}[1]{\textcolor{blue}{#1}}
\renewcommand{\newtext}[1]{#1}

\newcommand{\raunit}{\ensuremath{\mathrm{m}\Omega\cdot\mathrm{\mu}\mathrm{m}^2}}
\newcommand{\celcius}[1]{\ensuremath{#1\,^{\circ}\mathrm{C}}}
\begin{document}
\title{Interface characterization of Co${_2}$MnGe/Rh${_2}$CuSn Heusler
  multilayers} \author{Ronny Knut} \affiliation{Department of Physics
  and Astronomy, Uppsala Universtiy, Box 516, 75120 Uppsala, Sweden}
\author{Peter Svedlindh} \affiliation{Solid
  State Physics, Department of Engineering Sciences, Uppsala
  University} \author{Oleg Mryasov} \affiliation{Department of Physics
  and Astronomy and MINT Center, University of Alabama, Tuscaloosa}
   \author{Klas Gunnarsson} \affiliation{Solid
  State Physics, Department of Engineering Sciences, Uppsala
  University} 
\author{Peter Warnicke} \author{Dario Arena} \affiliation{National
  Synchrotron Light Source Brookhaven National Laboratory Upton, New
  York 11973, USA} \author{Matts Bj\"orck} \affiliation{Department of
  Physics and Astronomy, Uppsala Universtiy, Box 516, 75120 Uppsala,
  Sweden}\author{Andrew Dennison}\affiliation{Institut Laue-Langevin, Grenoble 38042, France} \author{Anindita Sahoo}\affiliation{Department of Physics,
  Indian Institute of Science, Bangalore-560 012, India} \affiliation{Solid State and Structural Chemistry Unit,
  Indian Institute of Science, Bangalore-560 012, India}  \author{Sumanta
  Mukherjee}  \author{D.D. Sarma} \affiliation{Solid State and Structural Chemistry Unit,
  Indian Institute of Science, Bangalore-560 012, India} \author{Sari
  Granroth} \affiliation{Department of Physics, University of Turku,
  Finland} \author{Mihaela Gorgoi} \affiliation{Helmholtz Zentrum
  Berlin f\"u{}r Materialien und Energie GmbH, BESSY II, Berlin,
  Germany} \author{Olof Karis} \affiliation{Department of Physics and
  Astronomy, Uppsala Universtiy, Box 516, 75120 Uppsala, Sweden}

\date{\today}

\begin{abstract}
  % Heusler alloys have been considered as promising materials for
  % different types of spintronic applications for several decades
  % owing to the high degree of spin-polarization predicted for the
  % ferromagnetic full-Heusler (X${_2}$YZ ) structure. Very recently
  % there has been a renewed interest in layered Heusler structures as
  % they have been proposed as a suitable source of spin current for
  % the next generation magnetic sensors relying on
  % current-perpendicular-to-plane giant magneto-resistance
  % (CPP-GMR). Indeed there has been several reports recently where
  % full-Heusler alloys have been used in devices resulting in
  % encouragingly high MR values, exceeding tens of percent in some
  % cases. The highest reported values are however reported for
  % conditions unsuitable for device fabrication and achievable
  % numbers seems to be much lower for conditions compatible with
  % fabrication parameters. It is generally thought that the low MR
  % values are associated with disorder in the Heusler structure
  % itself and interface properties in the layered device structure,
  % and routes to determine an optimized process is highly desirable.
  
  All-Heusler multilayer structures have been investigated by means of
  high kinetic x-ray photoelectron spectroscopy and x-ray magnetic
  circular dichroism, aiming \newtext{to} address the amount of
  disorder and interface diffusion induced by annealing \newtext{of
    the multilayer structure}. The studied multilayers consist of
  ferromagnetic $\mbox{Co}_2\mbox{MnGe}$ and non-magnetic Rh$_2$CuSn
  layers with varying thicknesses.  We find that diffusion begins
  already at comparably low temperatures between \celcius{200} and
  \celcius{250}, where Mn appears to be most prone
 to
  diffusion. We also find evidence for a 4 {\AA} thick magnetically dead
  layer that, together with the identified interlayer diffusion, are
  likely reasons for the small magnetoresistance 
  found for current-perpendicular-to-plane giant magneto-resistance
  devices based on this all-Heusler system.

  % To be filled out!
  % \begin{description}
  % \item[Background] This part would describe the context needed to
  %   understand what the paper is about.
  % \item[Purpose] This part would state the purpose of the present
  %   paper.
  % \item[Method] This part describe the methods used in the paper.
  % \item[Results] This part would summarize the results.
  % \item[Conclusions] This part would state the conclusions of the
  %   paper.
  % \end{description}
\end{abstract}

\maketitle
\section{Introduction}
Magnetic read head technology has experienced several large paradigm
changes during the last 15 years. Anisotropic magnetoresistance read
heads were surpassed by read heads based on giant magnetoresistance
(GMR) in 1996 enabled by the Nobel-awarded findings of Gr{\"u}nberg
and Fert \cite{BAIBICH:1988ys,BINASCH:1989vn}. Very soon after the
discovery of tunnel magneto-resistance (TMR), record values of the
magnetoresistance (MR) were reported for sensors based on this
effect. Today, TMR structures require a very thin MgO tunneling
barrier (typically $\sim 1$ nm) to achieve a sufficiently low
resistance of the device and further significant decrease of the
barrier thickness appears unrealistic. It has been suggested that
current-perpendicular-to-plane (CPP) GMR structures, as opposed to
current in-plane structures used in the first generation GMR devices,
are candidates to become the next generation of MR sensors as they do
not suffer resistance issues due to the all-metallic design
\cite{Takagishi2010}.

Heusler alloys are ternary alloys with the composition X${_2}$YZ,
where X and Y in general are two different transition metal atoms and
Z is a group 3 or 4, non-metallic element. A half-metallic character,
i.e. that the density of states of the majority band is metallic
while the minority band exhibits a gap at the Fermi level, was
theoretically predicted for the ferromagnetic (FM) half-Heusler alloy
NiMnSb \cite{deGroot_PRL_1985} followed by the extensive investigation
of properties originating from this remarkable feature of the
electronic structure \cite{Katsnelson_RevModPhys_2008}. This Heusler
alloy was investigated as a candidate material for a novel CPP-GMR
structure with Cu as spacer layer \cite{Caballero19981801}. Recently,
Co based full-Heusler alloys renewed the interest for this
topic. Experimental results for two types of non-magnetic (NM) spacers
have been reported fairly recently; (i) elemental metal spacers
\cite{APEX.2.063003, nakatani:212501} with (001) texture and (ii)
non-magnetic Heusler alloy spacers \cite{Ambrose_Mryasov,
  Ambrose_Mryasov_Patent, Mryasov_APL_2009} with (110)
texture. Combinations of the (001) textured FM Heusler alloys
Co$_2$MnSi and Co$_2$Fe(Si$_{0.5}$Al$_{0.5}$) with Ag yielded MR
values in excess of 28\% and a change in the resistance-area product
($\Delta$RA) of 8.8 {\raunit}.  \cite{APEX.2.063003, nakatani:212501}
More recently, enhancement of CPP-GMR up to 70\% MR and a $\Delta$RA
of 16.8 {\raunit}
% $m\Omega\cdot\mu\mathrm{m}^2$
have been measured for Co$_2$(Fe$_{0.4}$Mn$_{0.6}$)Si electrodes with
(001) Ag as spacer. \cite{APEX.4.113005}

The half-metallic character has been theoretically predicted for ideal
$\mbox{L2}_1$ crystals. For \newtext{actual samples} a certain amount
of disorder is expected, which will have a negative impact on the
value of the spin polarization
\cite{Carey_APL_2004,Ozdogan:2011qy,Galanakis:2009rt,Ravel20022812,Lezaic:2011fj}. The
CPP-GMR in all-Heusler structures can be affected by different types
of disorder, including bulk FM, bulk NM disorder and disorder specific
to FM/NM interfaces. Ambrose and Mryasov \cite{Ambrose_Mryasov_Patent,
  Ambrose_Mryasov} proposed a combination of FM and NM Heusler alloys
to maximize the spin asymmetry at the FM/NM interface. They argue that
a structure with the appropriate combination of FM and NM Heusler
alloys could provide a large, spin dependent, interface contribution
to the magnetoresistance \cite{Ambrose_Mryasov_Patent,
  Ambrose_Mryasov}, which, due to the non-Stoner like spin-splitting
in the FM Heusler will be less sensitive to disorder. \cite{Mryasov_APL_2009}
 
For device fabrication, due to compatibility with existing processes,
it would be advantageous to use the (110) textured combination of
Co$_2$MnGe (CMG) and the non-magnetic Heusler alloy Rh$_2$CuSn (RCS)
\cite{Ambrose_Mryasov, Ambrose_Mryasov_Patent}. A prototype hard
disk drive reader with a MR of about 7\% and a $\Delta$RA of about 4.0
{\raunit} has been constructed by Nikolaev et al.~
\cite{Mryasov_APL_2009} using such considerations.  % Nikolaev et
% al.~\cite{Mryasov_APL_2009} fabricated an all-Heusler CPP-GMR
% structure according to this scheme using RCS and CMG as the NM and
% FM layers, respectively. An MR of 6.7\% at RT was achieved, which is
% higher than what has been reported for CPP-GMR devices based on
% normal FM--NM transition metal structures.
However, it is still much lower than for TMR structures and too low
for being a viable alternative in sensor technology where $\mbox{MR}$
ratios in the range of 50\% at $\Delta{}\mathrm{RA}$ products in the
order of $0.1\, \Omega\cdot{}{\mu}\mathrm{m}^{2}$ are considered to be
required in the roadmap for magnetic media beyond
$2\mbox{TBit}/\mbox{inch}^2\,$. \cite{Takagishi2010}

In this work, we have studied the FM/NM interfaces in all-Heusler
multilayer structures. % by means of high kinetic x-ray photoemissyion
% spectroscopy (HIKE, also commonly referred to as HAXPES for Hard
% X-ray photoelectron spectroscopy) and x-ray magnetic circular
% dichroism measurements (XMCD)
The work focuses on changes induced by post-growth low temperature
annealing and it is shown that diffusion of atomic species is
initiated already in the temperature range between \celcius{200} and
\celcius{250}.

\section{Experiment}
We have studied multilayer samples comprised of the full-Heusler
compound CMG as the magnetic layer between non-magnetic layers of the
full-Heusler RCS.  The samples were grown using a commercial magnetron
sputtering system (Canon Anelva C7100) \cite{Mryasov_APL_2009}. The
aim of this study is to understand the effects of post-growth
annealing on interface quality and how these depend on the individual
layer thicknesses. The magnetic properties and in particular the
quality of the interfaces are studied using x-ray magnetic circular
dichroism (XMCD) \cite{PhysRevLett.70.694,PhysRevLett.68.1943},
\newtext{polarized neutron reflectivity} and SQUID magnetometry. The
modifications of the interfaces were characterized by
means of hard x-ray photoelectron spectroscopy (HAXPES or HX-PES, also
commonly referred to as HIKE for high kinetic energy photoelectron
spectroscopy)
\cite{holmstrom2006,granroth:094104,unp-go.sv.ea:08}. Additional
characterizations using ferromagnetic resonance \newtext{and} x-ray
resonant magnetic scattering \todelete{and polarized neutron
  reflectivity} have also been performed and will be published
elsewhere \cite{Ronny_Heusler_Unpub}. The XMCD experiments were
performed at beamlines I1011 and D1011 at the synchrotron facility
MAX-lab in Lund, Sweden, with 90\% and 75\% circularly polarized light,
respectively. All data were obtained using total electron yield. The
samples were magnetized in-plane after which they were measured in
remanence at room temperature. The HAXPES measurements were conducted
using the HIKE station \cite{unp-go.sv.ea:08} at the KMC-1 beamline of the BESSY II
synchrotron facility at Helmholtz Zentrum Berlin, Germany.  
The samples were heated to different temperatures in the
range \celcius{200} - \celcius{500} at a constant rate of
\celcius{10}/min and kept at constant temperature for 10 min, after
which the samples were cooled down to room temperature (RT) before a
measurement.

\newtext{Polarized neutron reflectivity were conducted up to the first
  Bragg peak at the SuperAdam instrument at the Institut
  Laue-Langevin. The instrument was operated with a highly oriented
 pyrolytic graphite monochromator and two Bragg mirrors to polarize the
  incident beam, producing an incident wavelength of 4.4 \AA. The
  reflected beam was analysed with a supermirror.  After each
  measurement the sample was annealed at the next temperature for 1 h
  under vacuum and was subsequently allowed to oven cool during half a
  day. Note that the heating protocol used for the neutron
  measurements is different to the one used for the other measurements
  primarily due to the available hardware at the different
  facilities. As diffusion is found to be present already at the
  lowest temperatures used (\celcius{100}), we expect that the neutron
  data are representative of a further progressed diffusion process,
  when compared to data obtained in the other measurements for the
  same annealing temperature.}  In Table \ref{tab:Samples}, we list the
samples used in this study. The samples have different combinations of
CMG and RCS layers thicknesses. Three of the samples are given
descriptive names since these are the most extensively studied, while
the other samples will be explicitly denoted when discussed.
\begin{table}
  \caption{\label{tab:Samples} Sample annotations and thicknesses of CMG and RCS layers.}
  \begin{ruledtabular}
    \begin{tabular}{ccccc} %options:c,d,...
      Annotation & CMG (\AA) & RCS (\AA)\\
      \hline
      Thick non-magnetic layer & 6 & 18\\
      Thick magnetic layer & 18 & 6\\
      Thick layers & 18 & 18\\
      - & 24 & 18\\
      - & 18 & 12\\
      - & 18 & 12\\
    \end{tabular}
  \end{ruledtabular}
\end{table}
 
\section{Results}
\begin{figure}
  \begin{center}
    \includegraphics[width=0.9\columnwidth]{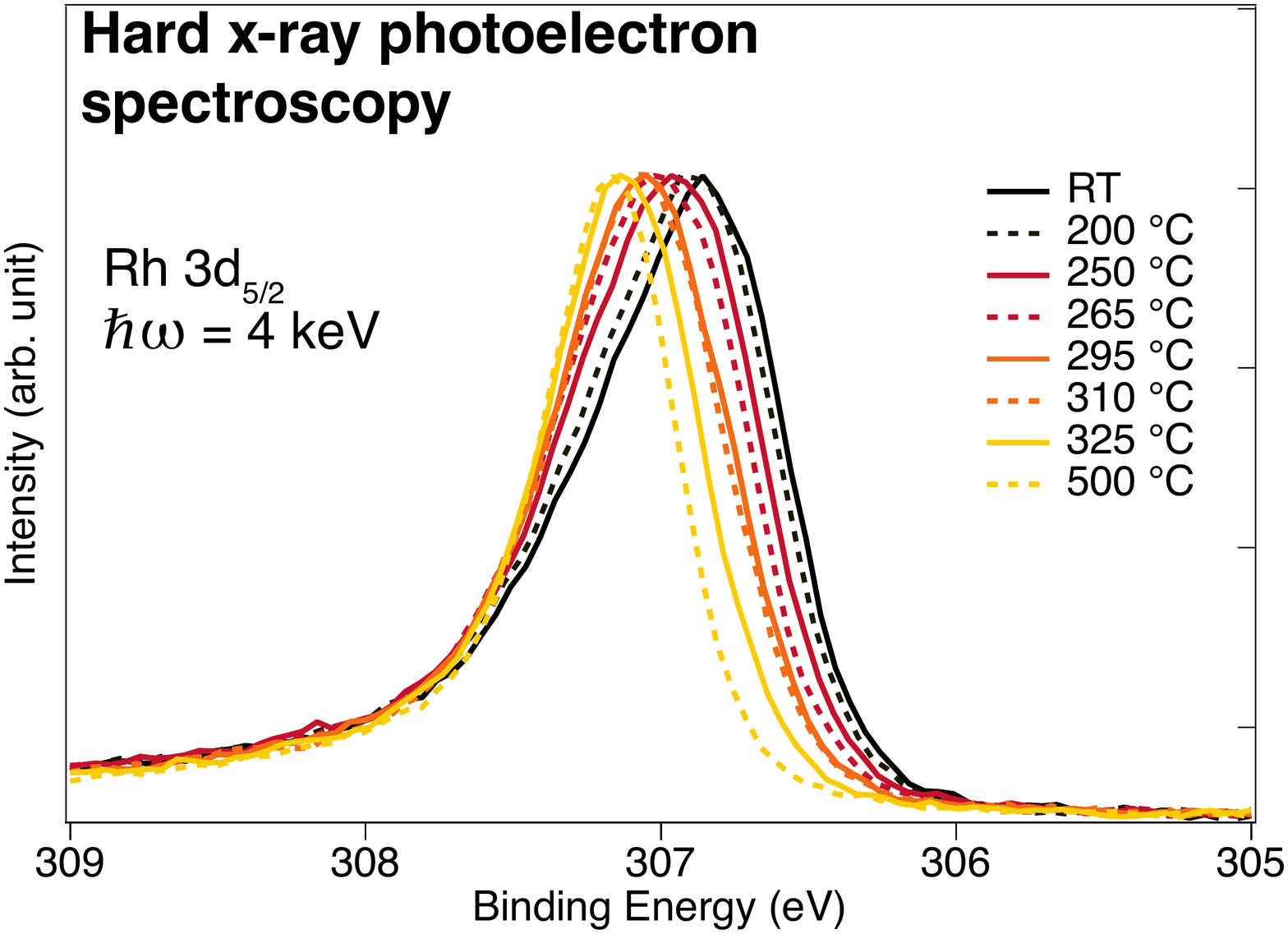}
  \end{center}
  \caption{\label{Rh_HIKE_PES} (Color online) Photoemission of Rh
    5d$_{5/2}$ from sample CMG 18\AA/RCS 18{\AA} (hereafter denoted
    'Thick layers') after different annealing temperatures. There is a
    peak shift to higher binding energy with increasing annealing
    temperature. The spectrum obtained  for \celcius{265} (red
    dash-dotted line) has been deconvoluted in
    Fig.\ \ref{Rh_fit}. }
\end{figure}
Regularly, beamlines dedicated to photoemission seldom offer photon
energies above 1 keV with reasonable photon intensity and resolution,
which limits the electron mean free path to about 5 {\AA}.  The KMC-1
beamline at HZB delivers 2 - 12 keV photons with high resolution,
making it ideal for studying bulk properties
\cite{unp-go.sv.ea:08,granroth:094104,holmstrom2006}. All the
photoemission results described here have been obtained using a photon
energy of 4 keV giving an estimated electron mean free path of 50 \AA.

As an atom in a solid is ionized by photons, the system will adapt to
the presence of the positively charged ion by screening of the
same. In general this involves both inter-atomic screening and
intra-atomic relaxation. In metallic systems, a substantial fraction
of the screening is due to mobile conduction electrons. The
effectiveness of the screening will affect the kinetic energy of the
photoelectron and hence also the measured binding energy of that
electron. The screening of the ionized atom depends on the nature of
electronic structure of the local environment and the hybridization
between the ionized atom and its surrounding. This is known as the
chemical shift in core level spectra \cite{Johansson:1980fk}. We use
this fact to distinguish between bulk and interface atoms and study
how these are affected by annealing.
 
\begin{figure}
    \begin{center}
      \includegraphics[width=0.9\columnwidth]{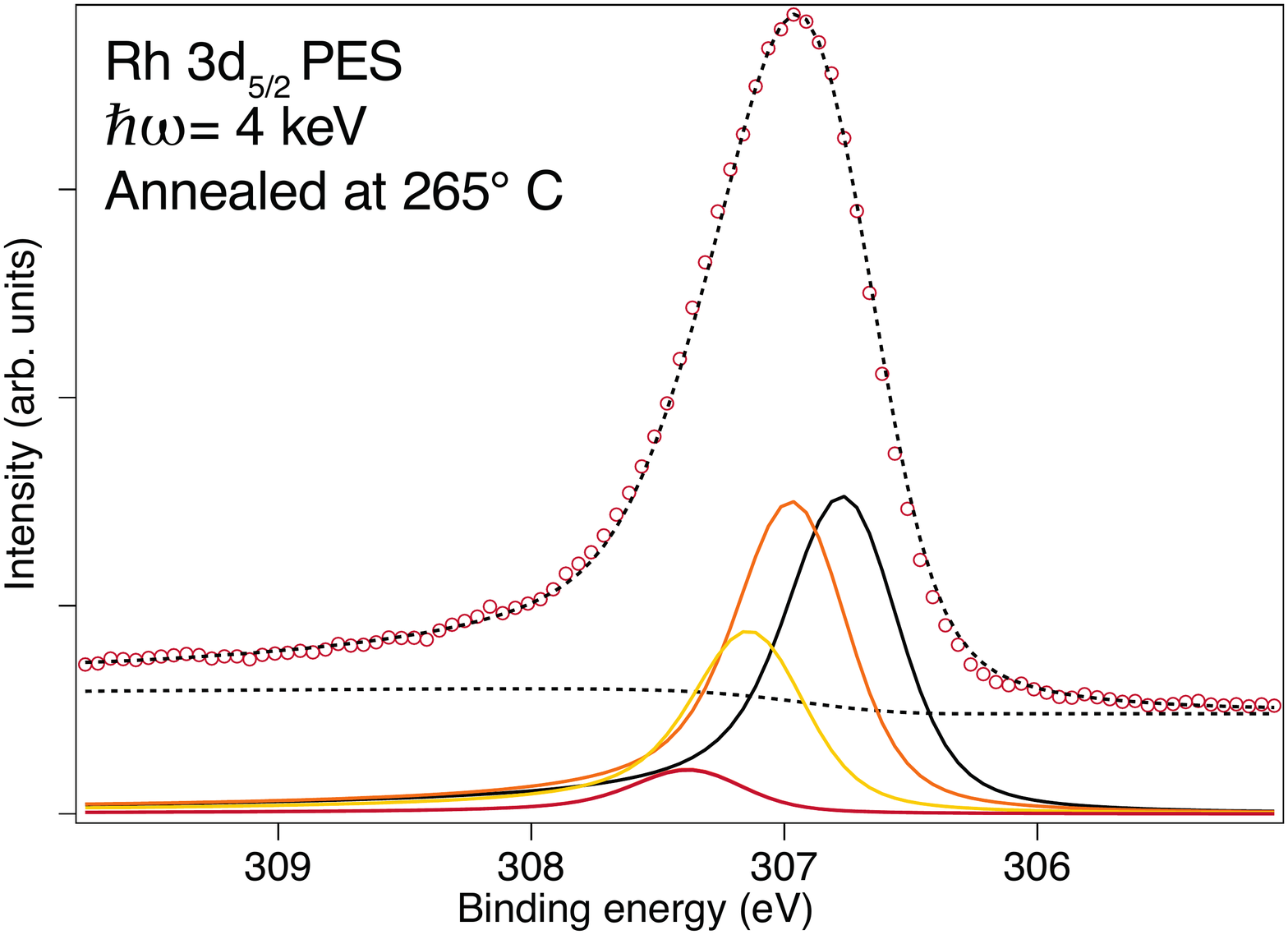}
    \end{center}
    \caption{\label{Rh_fit} (Color online) Deconvoluted spectrum of
      the Rh 3d$_{5/2}$ core level, fitted with 3 main peaks and a
      peak at the high BE shoulder, which is always 14\% of the two
      main peaks at the lower BE side.}
  \end{figure}

The sample with the thick CMG and RCS layers has been studied for
several annealing temperatures between \celcius{200} and
\celcius{500}, while the thick non-magnetic layer and thick magnetic
layer samples have been measured for a few selected annealing
temperatures. The Rh 5d$_{5/2}$ photoemission core level is shown in
Fig.~\ref{Rh_HIKE_PES} for the sample denoted 'Thick layers'
(conf.~Table~\ref{tab:Samples}). The peak clearly moves to higher
binding energy (BE) with increasing annealing temperature. It is
relatively speaking easy to acquire this core level with good
statistics. In combination with the large chemical shifts found for this Rh core level, makes it
possible to reliably fit the spectra with multiple peaks, as illustrated in
Fig.~\ref{Rh_fit}.  All lineshapes are set to Doniac-Sunjic type
\cite{Doniach:1970fk}, \todelete{convoluted with a fixed-width
  Gaussian representing essentially the instrumental broadening.}
\newtext{where the asymmetry and peak broadening are kept constant for
  every core level and chosen so that the experimental BE shift can be
  well described with as few peaks as possible. A Shirley-type
  background was used in the fit. \cite{Vegh:2006fk}}  

Each spectrum is
fitted with three main peaks; one component at 306.78 eV, one at
306.98 eV, and one at 307.14 eV binding energy, respectively. 
%The
% intensity of these peaks as a function of temperature are plotted in
% Fig.~\ref{Rh_theory}.
% The gaussian broadening is 380 meV which is what we expect from this
% beamline and the Lorenzian broadening is between 180-200 meV.
A fourth peak at high BE (307.36 eV) appears to be related to the two main peaks
at the lower BE side, since its intensity is always 14\% of these
peaks. 
  % The two peaks with lowest BE appear to have some kind of
  % 'satellite' structure (red) with the same BE (307.36 eV), we call
  % it a satellite since the intensity is proportional (14\%) to the
  % intensity of the main peaks.
  A common disorder in the full X$_{2}$YZ Heusler is X antisites in
  the Y atomic position. Thus, Rh anti-sites is a plausible
  explanation for this high BE peak, since it is not unlikely that
  14\% disorder of this type can occur \cite{Varaprasad_APEx_2010}.
  As the Rh begins to migrate due to heating, the disorder is cured at
  the same rate.
 
  \begin{figure}
    \begin{center}
      \includegraphics[width=0.9\columnwidth]{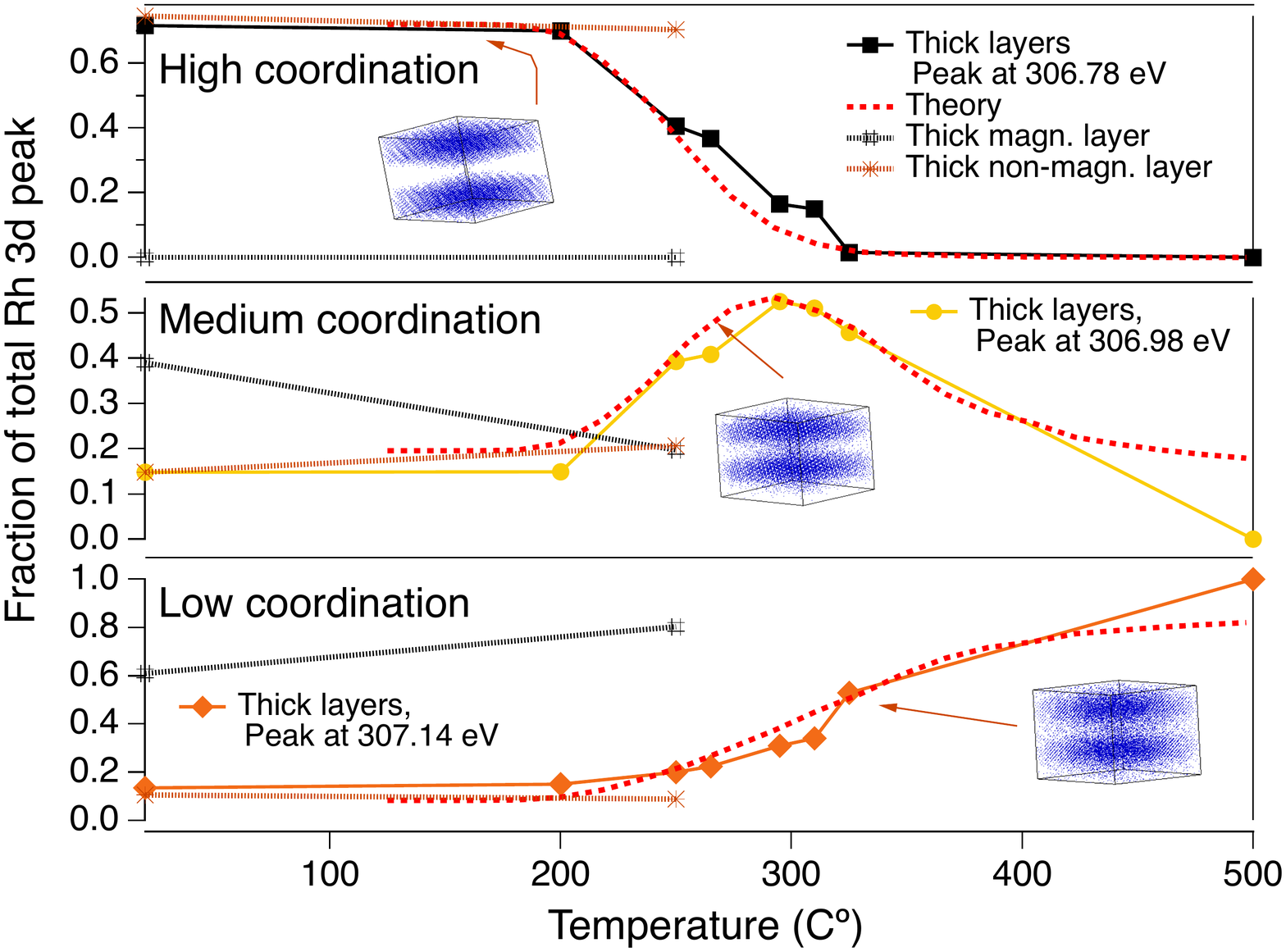}
    \end{center}
      \caption{\label{Rh_theory} (Color online) Fractions of the
        total peak areas of the  Rh $3 d_{5/2}$ core level spectrum, corresponding to the three main peaks
        obtained by fitting as illustrated in Fig.~\ref{Rh_fit}, are
        plotted as a function of annealing temperature. Top panel:
        high coordination component; middle panel: medium coordination
        component; bottom panel: low coordination component. Data for the
        three different samples considered are identified as indicated
        in the legends in each sub-panel. The dashed
        lines have been obtained from Monte Carlo simulations
        describing the 
variation of the high, medium and low coordination number cases with
annealing temperature.
% deviation from a perfect multilayer structure
%         due to disorder.
The three lattice illustrations show the Rh distribution for different
annealing temperatures. % Data for the 'thick magnetic layer' 
% (black dashed line with markers) and 'thick non-magnetic layer' (red
% dashed line with markers) samples are also given for RT and
% \celcius{250}, respectively.
}
\end{figure}

The intensities of the three main peaks a function of temperature,
obtained by spectral deconvolution as described above as, are plotted
in Fig.~\ref{Rh_theory}.  To better describe and understand the origin
of these three peaks we performed Monte~%
Carlo simulations, where the change of the coordination number, i.e.\
the number of Rh nearest neighbors (NN), was studied when introducing
random disorder as a consequence of heat treatment. The BE shift is
most sensitive to the number of Y and Z atoms Rh has as NN, and not to
the Rh NN, which are actually the next nearest neighbors. We made the
assumption that the composition of NNs is reflected in the number of
Rh neighbors. In reality, the Rh atoms occupy two fcc lattices in this
Heusler compound, which corresponds to a simple cubic structure with 6
NN. This structure did not give a good fit to the experimental results
since the number of Rh NN is too small to encompass all combinations
of realistic NN configurations. Instead we considered the next nearest neighbors and assumed that the Rh were
coordinated to 12 Rh neighbors, i.e.\ a local fcc-coordination of Rh.

A $26 \times 26 \times 26$ lattice was constructed with two layers of
Rh atoms, each 12 monolayers (ML) thick, representing the RCS
layers. The diffusion of Rh atoms was simulated by generating random
numbers (in the range 0 - 1) that should be smaller than
$\exp{\left[-\frac{E}{k_BT}\right]}$, where $E$ is an energy barrier,
for an atom to be allowed to move. It was enough to give every atom the
opportunity to move 600 times at each temperature, since the
simulation results became relatively insensitive to additional moves
beyond this number. The energy scale (i.e.\ reduced temperature) of
the simulation was scaled to the experimental results. In a perfectly
layered structure the amount of low coordinated atoms should be zero.
To obtain a more realistic starting configuration, a roughness
corresponding to a displacement of randomly chosen Rh atoms at the
interface, was introduced. To obtain agreement with experiments, we
found that, on average, every fourth interface Rh atom had to be moved.
% Therefore, a roughness was added by randomly moving 25\% of the
% atoms at an interface between CMG and RCS to the adjacent layer.
Illustrations of how the multilayer composition according to the
simulations changes with increasing annealing temperature have been
included as insets in Fig.~\ref{Rh_theory}. The fractions of atoms
with 12, 8-11 and 0-7 Rh neighbors are plotted as red dashed lines in
the top, middle and bottom graphs, respectively.
% 12 Rh neighbors is plotted as a red dashed line in top graph, the
% sum of 8-11 neighbors in the middle graph and the sum of 0-7
% neighbors in the bottom graph.  The results for the thick magnetic
% layer and thick non-magnetic layer samples correspond to RT and an
% annealing temperature of \celcius{250}.
Using a combination of statistical methods and analysis of core level
data we can thus obtain a qualitative and quantitative analysis of the
distribution of Rh as a function of temperature.
These results will be discussed further below, in connection to the
presentation of data from the other core levels.
% The thick non-magnetic layer sample is almost unaffected by heat
% treatment at 250 C and it is also very similar to the thick layers
% sample below 200 C. The thick magn. layer sample shows a sensitivity
% to heat treatment at 250 C.

\begin{figure}
  \begin{center}
    \includegraphics[width=0.9\columnwidth]{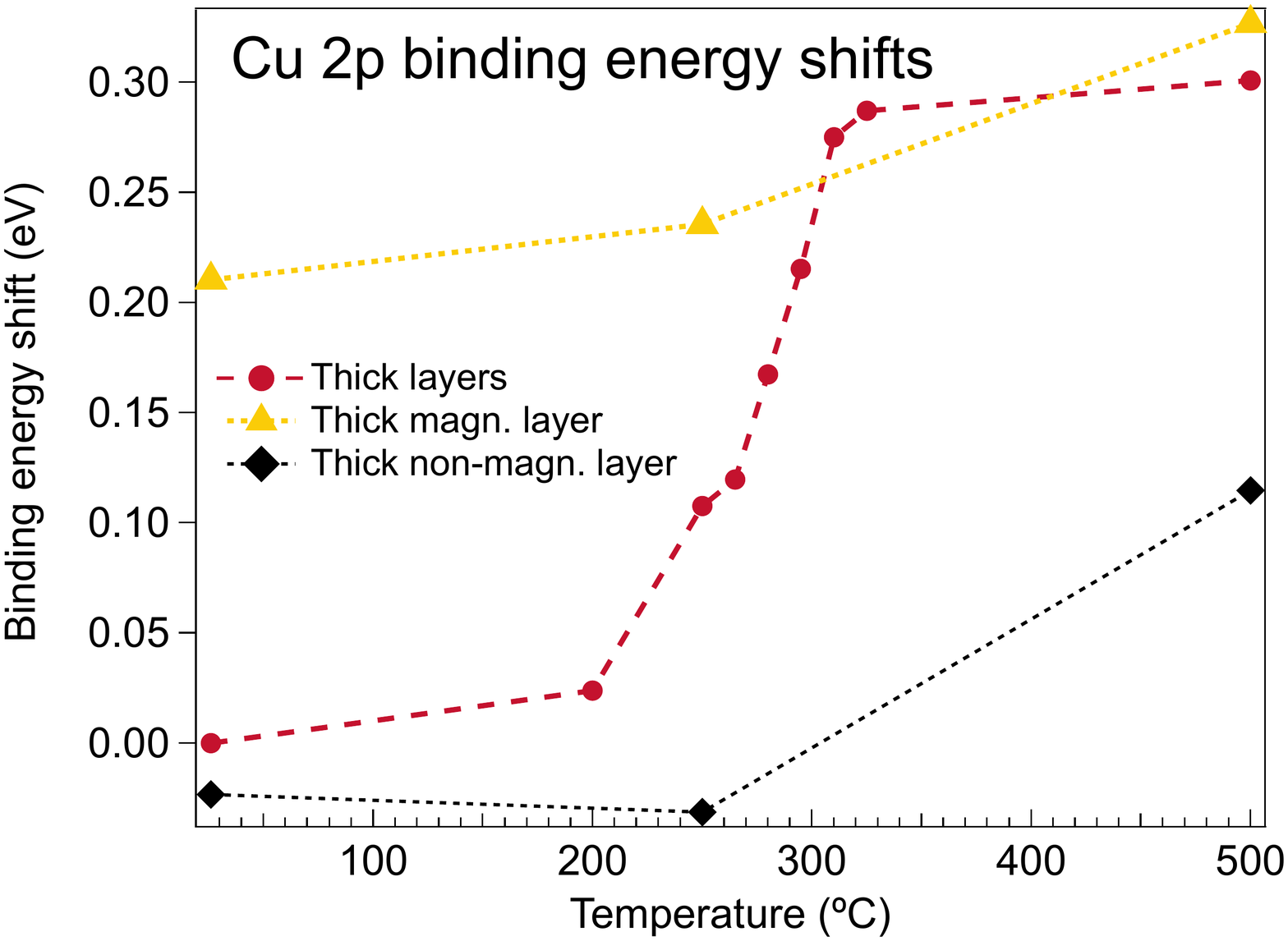}
  \end{center}
  \caption{\label{Cu_Shift_2} (Color online) The Cu 2p core level
    shift as a function of annealing temperature for three samples as
    indicated by the legends. The BE shifts  are given 
    relative to the position of the 2p level for the ``Thick layers''
    sample prior to annealing. Small shifts are found for samples with
    thick CMG layers ('Thick layers' and 'Thick magn. layer') already at \celcius{200}.
    % for the sample with thick RCS and CMG layers. 
    We observe no chemical shift below \celcius{250} for the sample
    with the thick non-magnetic layers, suggesting little modification
    below this temperature.}
\end{figure}

The Cu $2p$ core level spectra, obtained for different annealing
conditions, do not exhibit as large chemical shift as the Rh
$3d_{5/2}$ core level and the spectrum is always broader (not
shown). These factors makes it more difficult to use data for the Cu
$2p$ level for quantitative analysis like for Rh as described above. By
only considering the centroid of the spectrum fitted with a single
Doniac-Sunjic profile, we can however assign
chemical shift for each annealing temperature and sample. The result
is presented in Fig.\ \ref{Cu_Shift_2}. The core level shift is given
relative to the position of the $2p$ level for the ``Thick layers''
sample prior to annealing, which is thus set to zero. We observe that
the sample with thick magnetic CMG layers and thin RCS layers exhibits
a significant shift already before annealing. We therefore conclude
that data for both the Cu and Rh core levels
indicate that samples with a thin RCS layer have, as expected, a low
fraction of ordered bulk-like
RCS. In contrast we find that
the BE of the Cu $2p$ core level for the sample with the thick RCS
layer is very similar to the reference level of the ''Thick layers''
sample. 
% We also see that samples with a thin RCS layer
% appears modified already for an annealing temperature of \celcius{200}
% as judged from the Cu core level shifts.
Unlike the data for the Rh core level, the
Cu core level data thus seems to indicate a higher degree of order for
thinner magnetic layers.
The temperature dependence of the Cu $2p$ core level shift for all three samples suggests that
modifications occur already around \celcius{200}. Especially we find
that the core level shift of the ''Thick layers'' sample exhibit an
abrupt transition in the temperature range \celcius{200}-\celcius{300}.
% Cu also appears to be sensitive to heat treatment at 200 C.  This
% can be explained by a Rh deficient interface and will be discussed
% in the
% conclusions. % This also explains why a 12 monolayers thick Rh layer fitted the experimental data, while it should be closer to 7-8 monolayers.
% If the Rh closest to the interface has Cu and Sn neighbours it will
% have similar chemical shift as bulk Rh.

\begin{figure}
  \begin{center}
    \includegraphics[width=0.9\columnwidth]{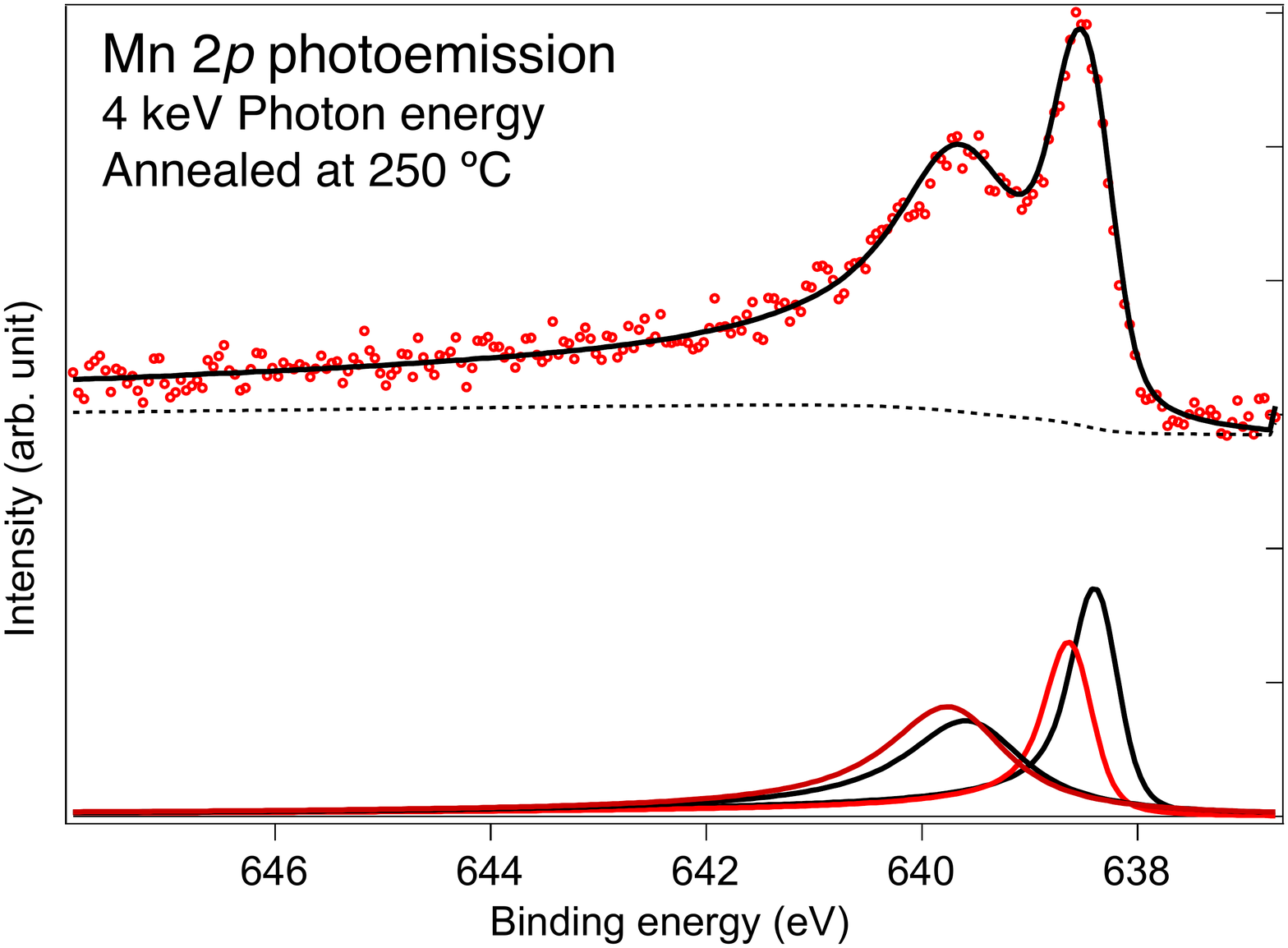}
  \end{center}
  \caption{\label{Mn_fit} (Color online) Deconvoluted Mn 2p spectrum
    recorded for the ``Thick Layers'' sample. The spectrum can be
    fitted by two components, and associated satellites, for all annealing temperatures. There is
    both a BE shift and a large difference in the satellite intensity
    between the components.}
\end{figure}

\begin{figure}
  \begin{center}
    \includegraphics[width=0.9\columnwidth]{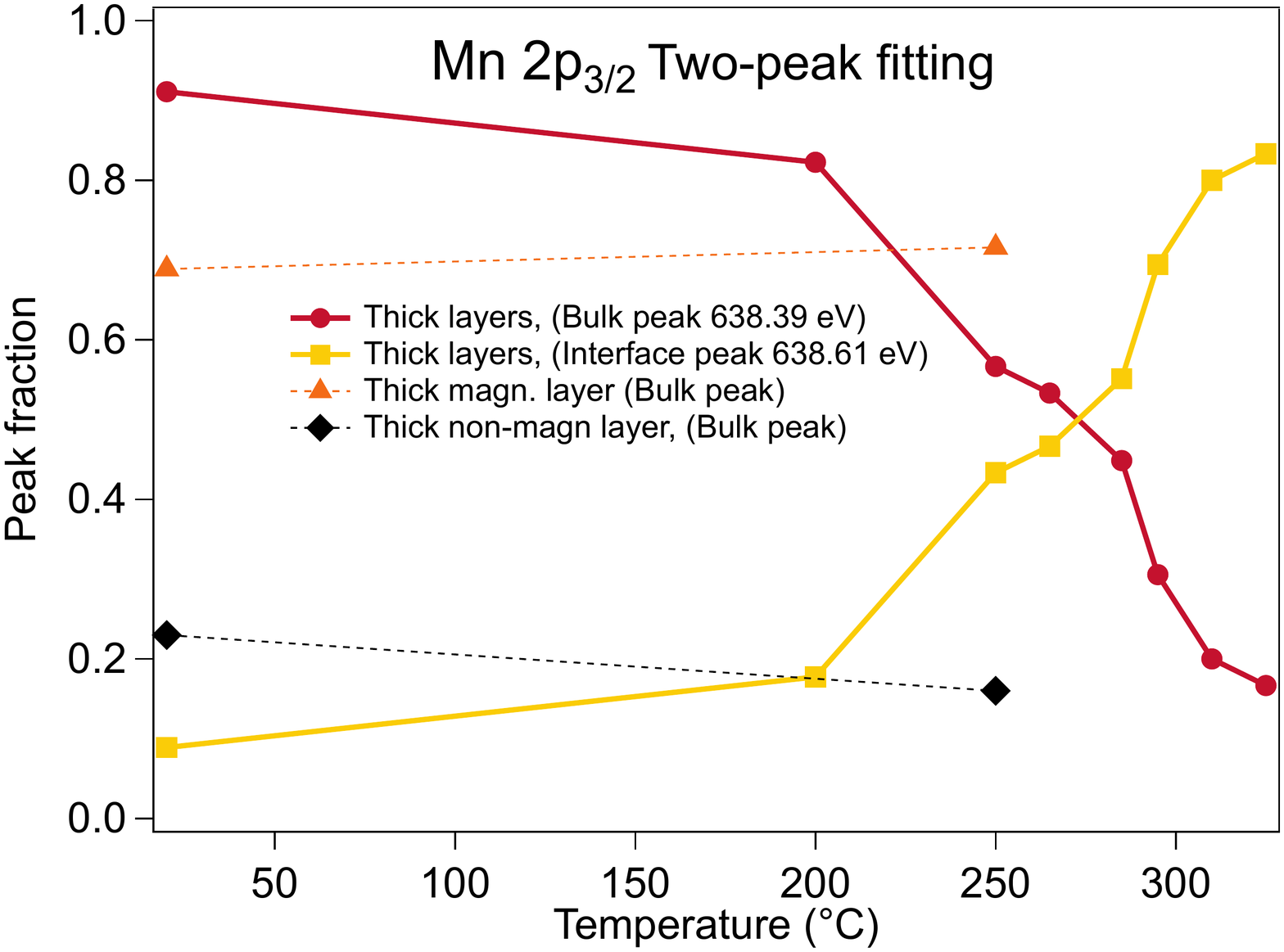}
  \end{center}
  \caption{\label{Mn_Shift} (Color online) The intensity of the bulk
    (circles) and interface (squares) peaks, derived from fits like
    illustrated in Fig.~\ref{Mn_fit}, are plotted as a function of
   annealing temperature for the 'Thick layers' sample. Only
    the bulk component is plotted for the other samples (triangles and
    diamonds).}
\end{figure}

The Mn 2p spectra have been fitted by two peaks that both exhibit
strong satellite structures, as illustrated in Fig.~\ref{Mn_fit},
where the results for the 'Thick layer' sample annealed at
\celcius{250} is shown. Also here, Shirley-type backgrounds were used.
In the fitting procedure, the intensity of the
satellite was kept constant relative to that of the main line. The
main peak at $\sim 638.4\,\mbox{eV}$ is attributed to the interior of
the layer (bulk peak), while the peak at $\sim 638.6\,\mbox{eV}$ is
attributed to the interface.  The
fraction of each peak as a function of temperature is shown in
Fig.~\ref{Mn_Shift}.  Similar to what was found for Cu above, there
are also differences in the temperature dependence of the core level
shift for the Mn $2p$ core level depending on the thickness of the
adjacent layer. Also, as for Cu 2p, we observe a changes for the Mn
$2p$ core level 
already for an annealing temperature of \celcius{200}. The
bulk component is smaller for the 'Thick magnetic layer' sample which
has a thinner RCS layer (triangles), which would suggest that the
interface quality is lower for thinner RCS layers. As expected, the
bulk component is also smaller for the 'Thick non-magnetic layer'
sample with thinner CMG layer (diamonds).

\begin{figure}
  \begin{center}
    \includegraphics[width=0.9\columnwidth]{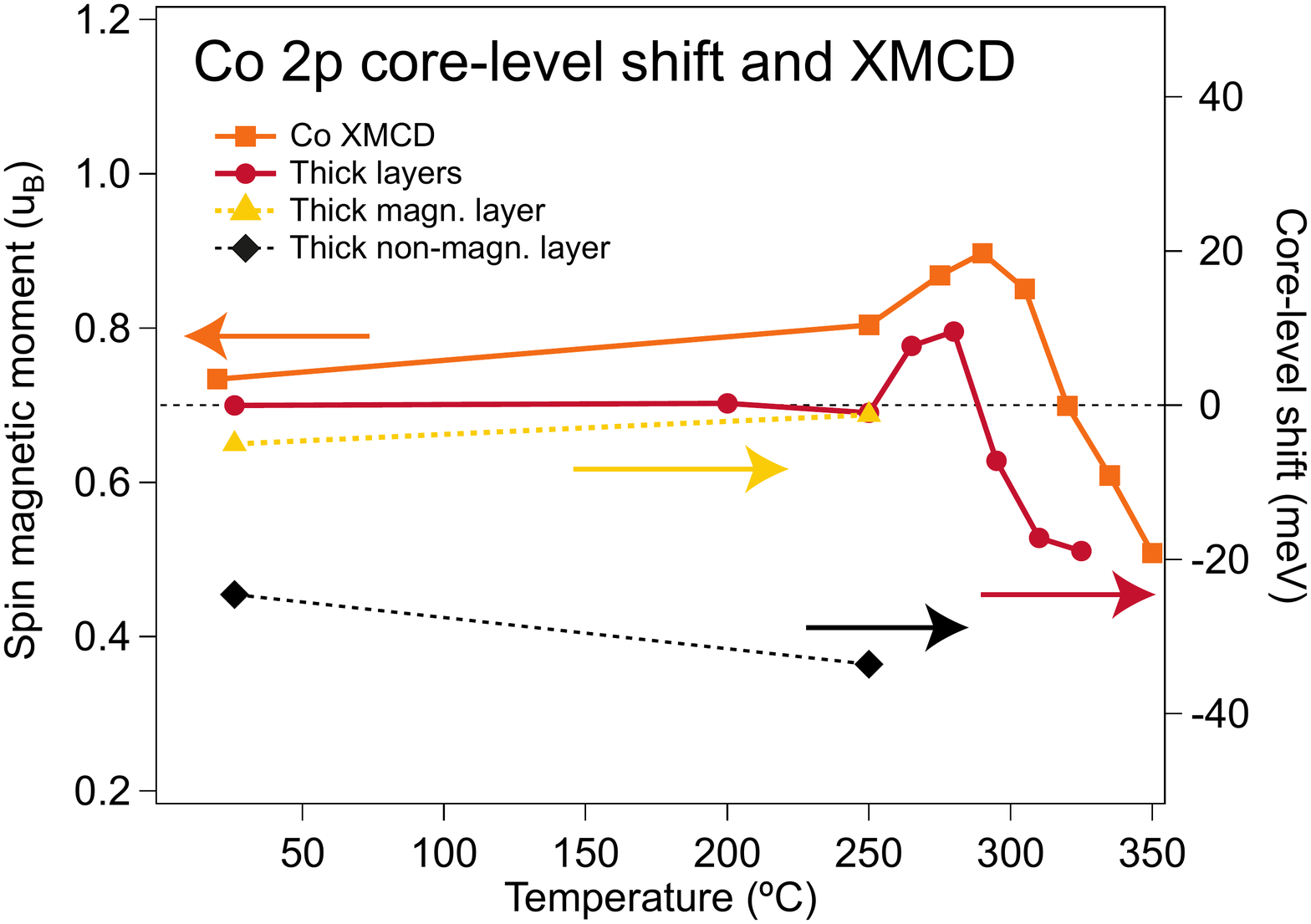}
  \end{center}
  \caption{\label{Co_Shift_XMCD} (Color online) The Co XMCD spin
    magnetic moment is plotted on the left scale. On the right scale
    we have plotted the chemical shift of the Co $2p_{3/2}$ core level. The behavior of the
    chemical shift and magnetic moment is very similar.}
\end{figure}

In Fig.\ \ref{Co_Shift_XMCD} we show both the chemical shift of the Co
$2p_{3/2}$ core level and the
spin magnetic moment of Co as obtained from an XMCD sum-rule analysis
\cite{PhysRevLett.70.694,PhysRevLett.68.1943}, as a function of
annealing temperature. As for Cu, the core level shift was obtained by
fitting a single Doniac-Sunjic line with a Shirley background to the data at each temperature
and sample. The samples with thick CMG layers, i.e.\ the
'Thick layers' and 'Thick magnetic layer', exhibit a
negligible chemical shift of the Co $2p_{3/2}$ core level up to \celcius{250}, suggesting little
modification at the X sites below this temperature. Between
\celcius{250} and \celcius{280}, we find an increasing positive
shift. For an annealing temperature of \celcius{300}, the chemical
shift becomes negative and we observe an increasing negative shift in
the interval \celcius{300} - \celcius{330}. The qualitative behavior
is very similar to the annealing temperature dependence of the spin
magnetic moment obtained from the XMCD measurements, which also shows an
increase in the same temperature range followed by a decrease. A
3d-hole count of 2.2 for Co and 4.2 for Mn\cite{Antonov_JAP_2006,
  PhysRevB.72.024437} was used for extracting the magnetic moment from
the XMCD results using the sum rules \cite{PhysRevLett.68.1943,
  PhysRevLett.70.694}. The formation of Co antisites, where Co atoms
occupy Mn sites, or Mn antisites, where Mn atoms occupy Co sites, are
expected to occur in the FM Heusler layers \cite{Mryasov_APL_2009,
  PhysRevB.69.094423, Ravel20022812}. According to theory, Mn
antisites will couple antiferromagnetically with its nearest neighbors, while for Co antisites
there is instead an increase of the ferromagnetic moment by 32\%. 
\cite{PhysRevB.69.094423, Ravel20022812} This could explain the
observed increase of the Co moment, since we know that Mn is mobile
already for an annealing temperature of \celcius{200} and it therefore
becomes possible for Co to occupy Mn sites. 
One may note that Co
antisites, unlike Mn antisites, are detrimental for the spin
polarization of the material \cite{PhysRevB.69.094423}.

\begin{figure*}
  \begin{center}
    \includegraphics[width=0.49\textwidth]{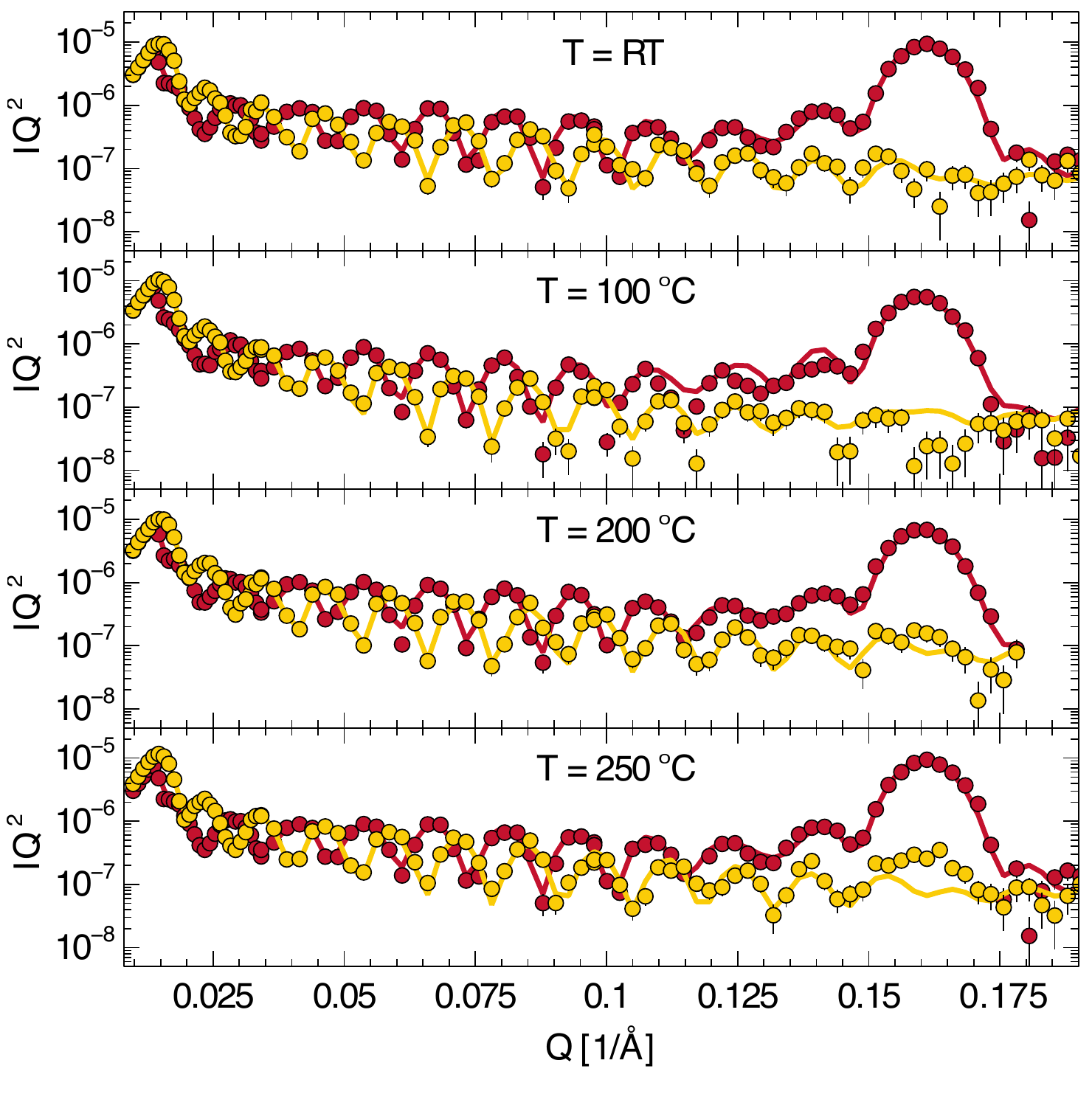}\hfill
    \includegraphics[width=0.49\textwidth]{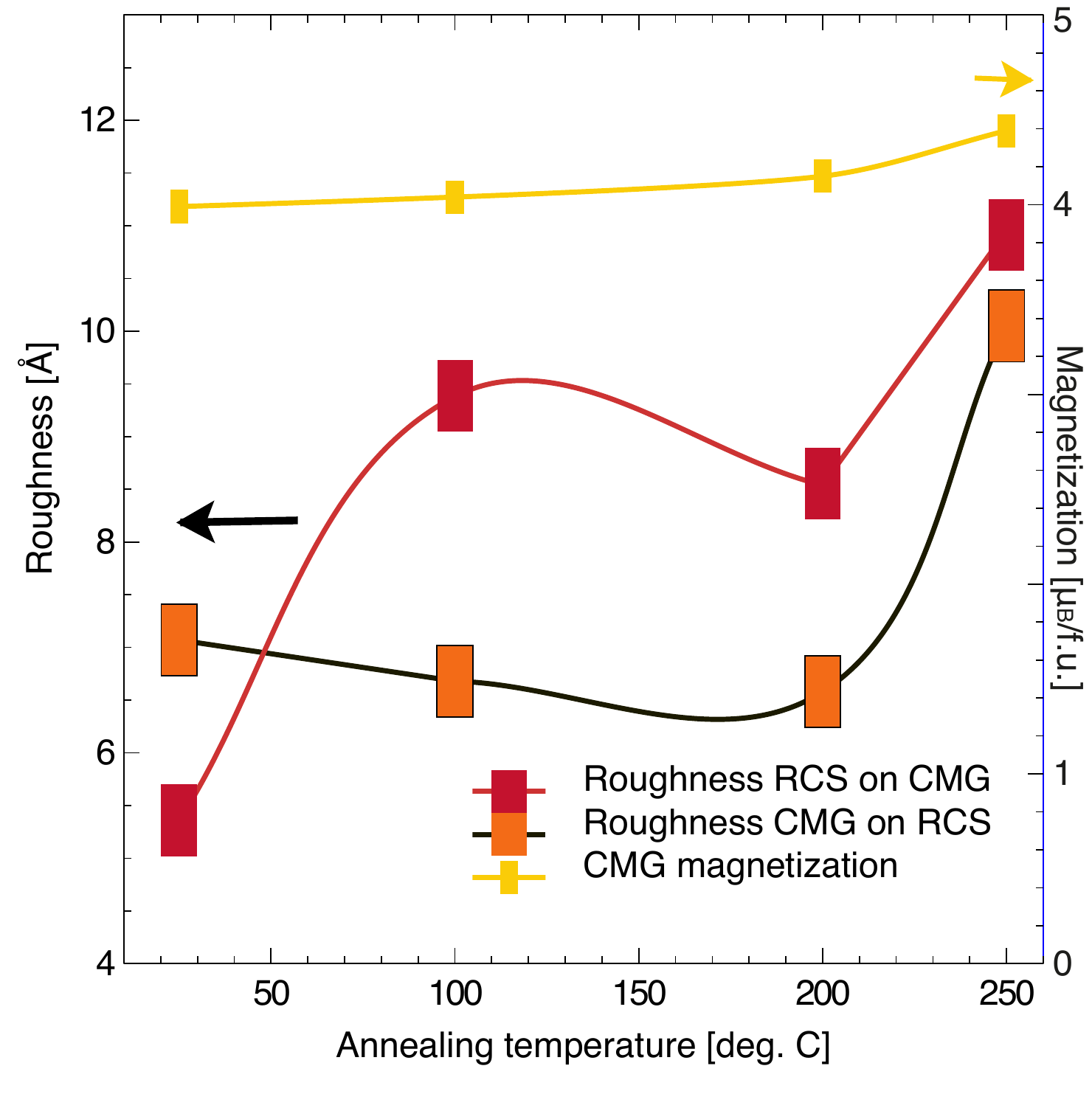}
  \end{center}
  \caption{\label{Neutron_data}\newtext{ (Color online) (left) Neutron
      reflectivity data where the scattering length density for
      the magnetic and nuclear scattering length are plotted with
      yellow and red colors, respectively. (right) The interface
      roughness and magnetic moment obtained from the refinement using
      GenX\cite{Bjorck20071174}. The roughness of RCS grown on CMG is
      very sensitive to annealing already at \celcius{100}. }}
\end{figure*}

\newtext{The sample consisting of 24 {\AA} CMG/ 18 {\AA} RCS was
  studied with neutron reflectivity to complement the spectroscopic methods. The sample was
  chosen due to its large layer thicknesses, which made it suitable
  for neutron reflectivity measurements. The data was analysed with
  the GenX program \cite{Bjorck20071174} by co-refining all data sets
  up to \celcius{300}, keeping the same 
  parameters for all temperatures except for
  the roughness of the RCS layer, CMG layer and the capping
  layer. In addition, the magnetic moment for the CMG layer was
  allowed to vary. For the four refined data sets, up to a temperature
  of \celcius{250}, there was a total of 51 free parameters yielding a
  fit with a crystallographic R1 factor of 5.7 \%. The results of the
  fits together with raw data can be seen in Fig.\ \ref{Neutron_data}
  (left). The resulting scattering length density (SLD) for the
  magnetic and nuclear scattering length are plotted with yellow and red
  colors, respectively. The substrate is located to the left in the
  plots and the film/air interface is located to the right. The
  parameters of interest in this study; the roughness of the
  multilayer interfaces and the magnetic moment of the CMG layer can
  be seen in Fig.\ \ref{Neutron_data} (right). It should be noted that
  after an anneal to \celcius{300} the multilayer Bragg peak has
  disappeared completely indicating a total intermixing of the layered
  structure (not shown). This temperature can consequently not be modelled by
  merely allowing the roughnesses to vary as there are no interfaces
  left, consequently it is left out of the discussion of the
  reflectivity data.  However, the large changes seen as the annealing
  temperature increases to \celcius{300} is in agreement with the HIKE
  data that also show layer interdiffusion at these
  temperatures. The first rise in the roughness of the RCS on CMG
  interface in Fig.\ \ref{Neutron_data} (right) is in accordance with
  the weak intermixing seen for the Mn and Cu core levels. The
  intermixing is mostly localized to one of the interfaces in the
  multilayer structure, RCS on CMG, and starts already at around a
  temperature of \celcius{100}. We again note the distinction between the
  protocols used for heating due to differences in the hardware used. The
longer annealing time used for the samples used in the neutron
measurements are likely giving rise to the apparent higher degree of
interdiffusion indicated by the neutron data.}
 
\begin{figure}
  \begin{center}
    \includegraphics[width=0.9\columnwidth]{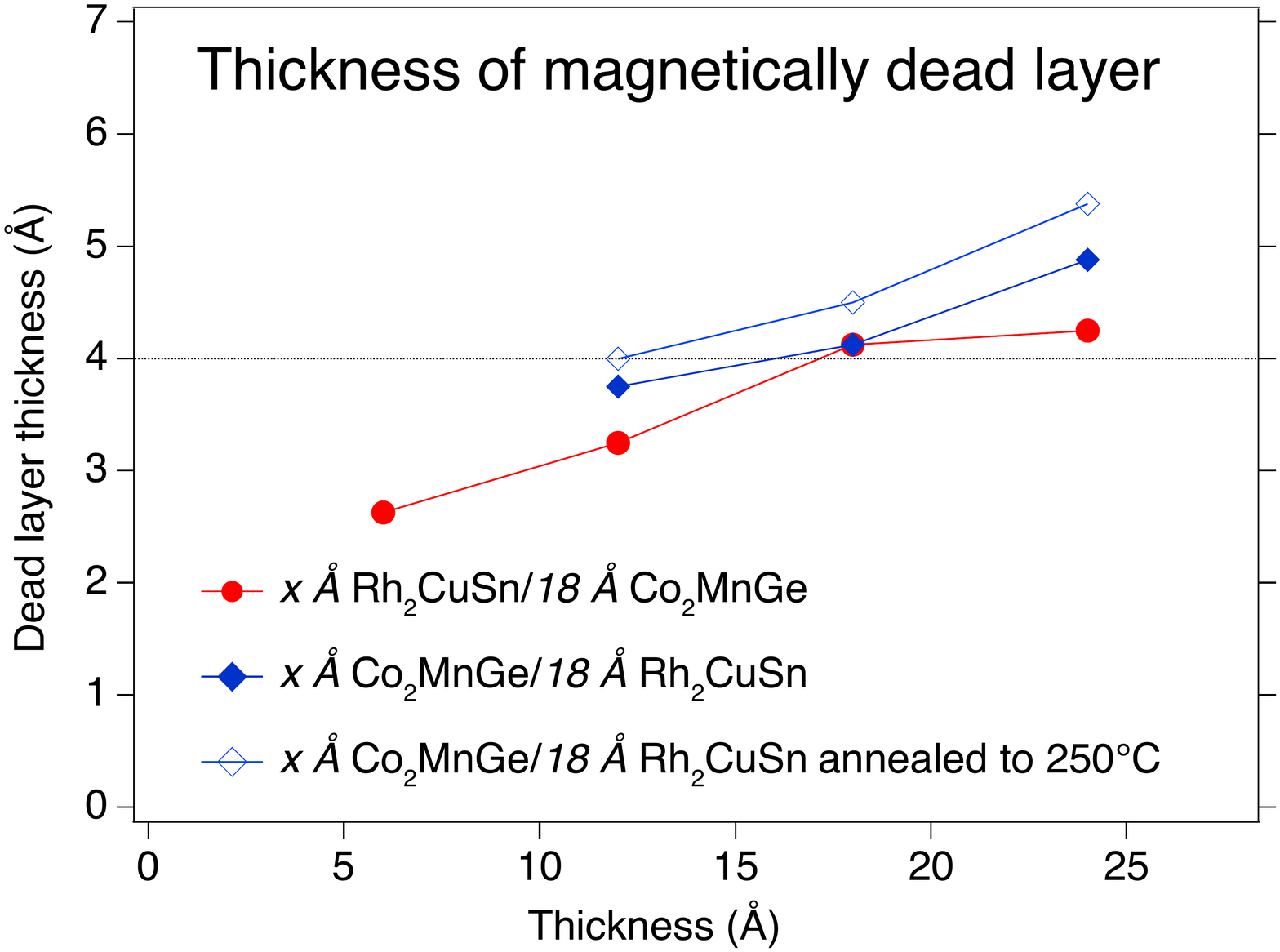}
  \end{center}
  \caption{\label{nr_dead_layers_new} \newtext{(Color online)
      Estimated thickness of the magnetically dead layer for different
      layer thicknesses obtained from SQUID magnetometry. (Red circles) The
      thickness for the CMG layer has been kept constant at 18 {\AA},
      while the RCS layer thickness is varied between 6-24 \AA. (Solid
      blue diamonds) The RCS layer thickness is 18 \AA{}, while the
      CMG thickness is varied between 12-24 \AA. (Unfilled blue
      diamonds) As previous data, but the samples have been annealed at
      \celcius{250}. The dead layer thickness is relatively
      insensitive to the thickness of the layers.}}
\end{figure}

\newtext{The saturation magnetic moments for all samples were obtained
  with SQUID magnetometry. The 6 \AA{} CMG/ 18 \AA{} RCS sample, which
  shows no remanence in XMCD, was used for providing the magnetization
  of the FeCo seed layer by assuming that only this layer contribute
  to the remanent magnetization obtained from the SQUID
  magnetometry. The magnetization of FeCo was found to be
  $1.5\cdot10^6$ A/m which is close to the bulk value of
  $1.74\cdot10^6$ A/m. 

  To account for the magnetization obtained by means of SQUID
  magnetometry we find that it necessary to include magnetically dead
  layers in each CMG layer. If one assumes the theoretical magnetic
  moment of 5 $\mu_B/f.u.$ in the CMG layer one finds that \newtext{a
    magnetically dead layer needs to be included, to account for the
    measured magnetic moment. The thickness of this magnetically dead
    will have to increase} from 5 to 10 \AA{} as the CMG layer
  thickness increases from 12 to 24 \AA. However, as illustrated in
  Fig.\ \ref{nr_dead_layers_new}, if one assumes a magnetic moment of
  4 $\mu_B/f.u.$ we find that the required thickness of the dead layer
  is much less dependent on the CMG/RCS thickness, with \newtext{an
    approximately constant} thickness of about 4 \AA. This corresponds
  well to the magnetic moment found by neutron reflectivity.% , which
  % however does not include a dead magnetic layer in the refinement
  % procedure
In Fig.\ \ref{nr_dead_layers_new}, the dead layer thickness for samples with 18 \AA{} CMG
  and varying RCS layer is plotted as red circles. For samples with
  18 \AA{} RCS and varying CMG layer, the dead layer thickness is plotted as blue diamonds. A
  small increase in the dead layer thickness is found for samples annealed to
  \celcius{250}.}

% neutron moment gŒr upp med lŒga T, Mn ordnar sig?.

% This very likely to occur due to curing of B$_2$ disorder where Co
% is located on a Y site. It has been showed that this disorder lowers
% the magnetic moment of the Co.  The Co moment is small compared to
% the expected 1.8 $\mu_B$ of metallic Co.  Fig. \ref{Co_XMCD_layers}
% illustrates why this is the case.

\begin{figure}
  \begin{center}
    \includegraphics[width=0.9\columnwidth]{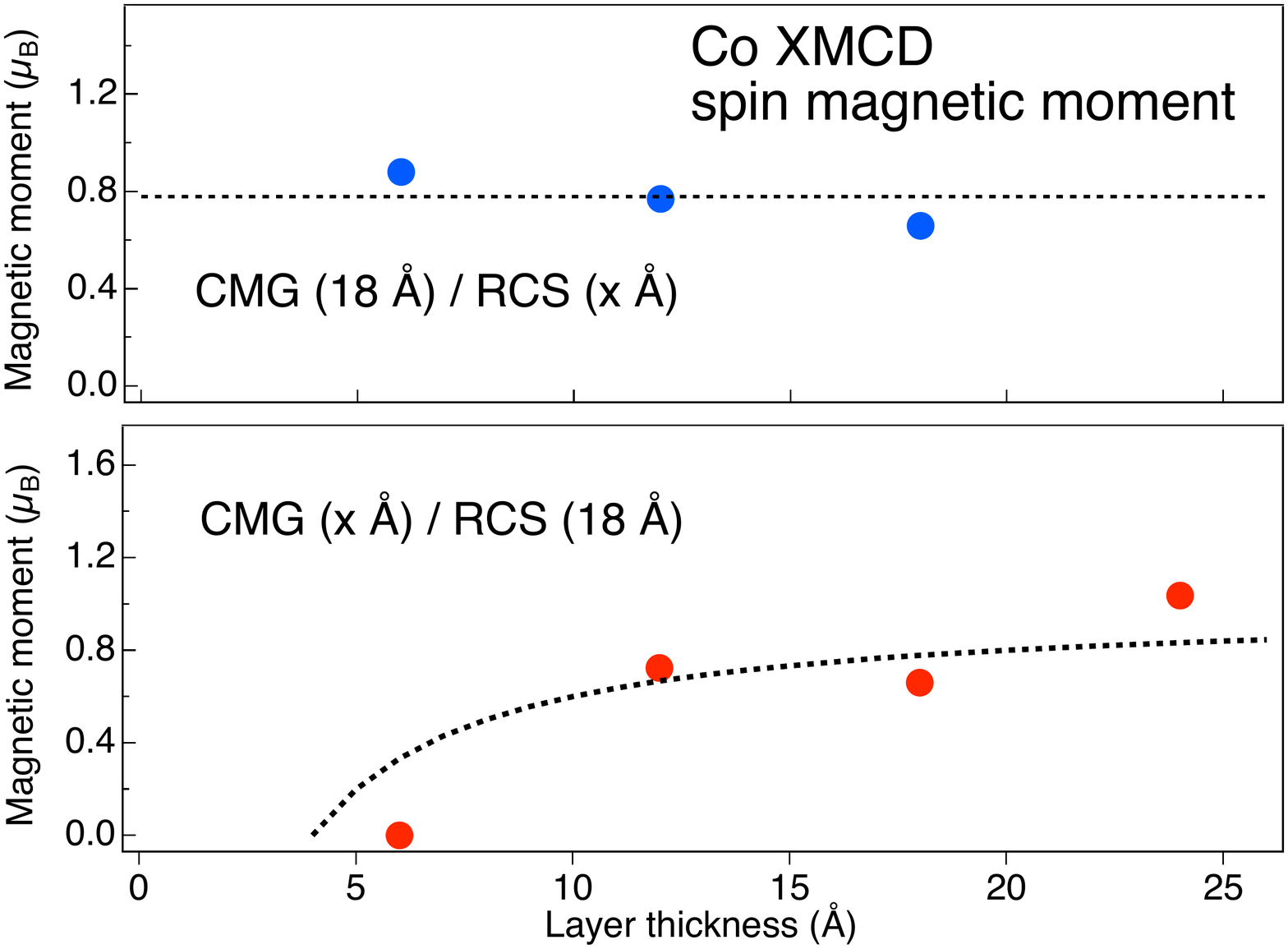}
  \end{center}
  \caption{\label{Co_XMCD_layers} (Color online) The Co spin magnetic
    moment for different CMG and RCS layer thicknesses before
    annealing. The top graph shows data for a constant CMG layer
    thickness of 18 \AA{} varying the RCS layer thickness, while the
    lower graph shows data for a constant RCS layer thickness of 18
    \AA{} varying the CMG layer thickness. The increase of the
    magnetic moment in the lower graph indicates a 4 {\AA} thick
    magnetically dead layer in the CMG layers.}
\end{figure}

The \newtext{XMCD} Co spin magnetic moment as a function of CMG and
RCS layer thicknesses are shown in Fig.~\ref{Co_XMCD_layers}.  In the
upper graph we give the data for a constant CMG layer thickness of 18
\AA{} varying the RCS layer thickness, while for the lower graph we
vary the CMG layer thickness keeping the RCS layer thickness constant
at 18 \AA{}.  
The black dashed line corresponds to the average Co
moment in the CMG layer, assuming a 4 {\AA} thick magnetically dead layer
and a Co spin moment of $1 \mu_B$
\cite{PhysRevB.82.184419,PhysRevB.81.144417}  in the remaining part of the CMG layer.
%
% \newtext{The black dashed line corresponds to the theoretical Co spin
%   magnetic moment of 1
%   $\mu_B$\cite{PhysRevB.82.184419,PhysRevB.81.144417} with a 4 \AA{}
%   dead magnetic layer.}
%
In the lower graph we observe an increase of the magnetic moment with
increasing CMG thickness. It should be explicitly noted that the 6
\AA{} CMG sample does not show any XMCD signal. Assuming a \newtext{4} \AA{} magnetically dead layer
in CMG and a Co magnetic moment of 1 $\mu_B$, we obtain a reasonable
fit to the experiment results as indicated by the black dashed line in
the lower graph of Fig.~\ref{Co_XMCD_layers}. 

%
% The black dashed line represents the magnetic moment if we assume a
% 5  magnetically dead layer in the CMG. Also, in this calculation
% the Co magnetic moment is 1.05 $\mu_B$. This is a reasonable
% assumption if we consider that the magnetic moment is decreased by
% an 18  RCS layer, see top graph of Fig. \ref{Co_XMCD_layers},
% from its initial value of 1.8 $\mu_B$.
%
% \comment{I don't understand the text below.}  The green square in
% the top panel of Fig.~\ref{Co_XMCD_layers} is the decreased Co
% moment, from 1.4 $\mu_B$, due to a 5 {\AA} magnetically dead layer
% with no weakening of the exchange interaction corresponding to a 0
% \AA{} thick RCS layer.
%
% The green square in the top graph is the calculated Co moment if we
% assume a 5  magnetically dead layer for a 18  thick CMG
% layer and no weakening of the exchange interaction.
The Mn XMCD is hampered by large background contributions but a
reliable spin magnetic moment could be derived for the sample with the
thick magnetic layer. Moreover, using the sum rules for Mn is not
straightforward due to mixing of the Mn $L_2$ and $L_3$ edges.
A correction factor of 1.47 has been proposed
\cite{JPSJ.65.1053}. However, Saito et al.\ \cite{PhysRevB.81.144417}
found that omitting the correction factor gives more reasonable values
of the Mn spin magnetic moment. Taking the 4 {\AA} magnetically dead
layer in consideration we obtain a Mn spin magnetic moment of \newtext{1.98} $\mu_B$ without considering a
correction factor. This gives a total magnetic moment of 3.93 $\mu_B$
in the unit cell, which is very close to the magnetic moment obtained
from neutron reflectivity. \todelete{experimental and theoretical
  value of 5 $\mu_B$. Using the correction factor we instead obtain
  3.13 $\mu_B$, which is closer to the theoretical value of 3 $\mu_B$
  for Mn. } \newtext{Others studies have found that the magnetic
  moment of Co increases for thin layers compared to bulk while the Mn
  is generally smaller
  \cite{PhysRevB.82.184419,PhysRevB.81.144417}. This can be explained
  by the disorder in thin films giving an increased magnetic moment
  for Co$_{\mathrm{Mn}}$ while Mn$_{\mathrm{Co}}$ is
  antiferromagnetically coupled to its nearest neighbors and will
  hence lower the saturation magnetization of the
  film. \cite{PhysRevB.69.094423}}
% obtains an AFM coupling and hence a
  % lower average magnetic moment.
 %are energetically very close and therefore a correcting factor of 1.47\cite{PhysRevB.82.184419} is used and the number of d-holes is calculated to be 4.5. The obtained spin magnetic moment is 2.37 $\mu_B$ and if also consider the dead magnetic layer we have 3.63 $\mu_B$.

\section{Conclusions}
\newtext{The core levels of Mn and Cu indicate weak intermixing of the
  CMG and RCS layers already for an annealing temperature of
  \celcius{200}, while for annealing temperatures above \celcius{250}
  we observe significant changes for all studied core-levels. We have shown that
  the Co atoms in the Co$_2$MnGe layer appear to be relatively
  immobile for annealing temperatures below \celcius{250}, while for
  higher annealing temperatures Co antisites are
  created. \todelete{This conclusion is evidenced by a combination of
    core level shifts and magnetic moment variations derived from XMCD
    results.}  These antisites have a negative impact on the spin
  polarization of the material. Varaprasad et
  al. \cite{Varaprasad_APEx_2010} showed that the amount of Co
  antisites can be reduced by exchanging 25\% of the Ge with
  Ga. \todelete{The remanent Co magnetic moment decreases for
    increasing RCS layer thicknesses, which can be explained either by
    a change of the interlayer coupling or a change of the CMG layer
    quality.} \comment{We might need to have to revise this conclusion
    based on SQUID data.} The increase of the Co magnetic moment with
  increasing CMG layer thickness can be explained reasonably well by
  assuming a 4 \AA{} magnetically dead layer in CMG. The total
  magnetization of CMG is about 4 $\mu_B$ which is lower than the bulk
  value of 5 $\mu_B$. This is likely due to $\mbox{Mn}_{\mathrm{Co}}$
  antisites which couple anti-ferromagnetically to their nearest neighbors, while
  $\mbox{Co}_\mathrm{Mn}$ antisites would give an increase in magnetic
  moment. \todelete{Mn exhibits more disorder and Cu a more ordered
    state for thinner adjacent layers.} Both Co and Rh, i.e.\ the A
  site atoms in two different layers, appear insensitive to the
  thickness of the adjacent layer before annealing.  \comment{Don't
    get the conclusion in the previous sentence.} Rh has about 10\% of
  the atoms in a low coordinated state before heat treatment which is
  in accordance with neutron reflectivity data suggesting a roughness
  of 6 \AA{} at the interfaces. Furthermore, the interface where
  RCS is grown on CMG appears to be sensitive to heat treatments below
  \celcius{250}. Our findings has profound consequences for
  applications based on these materials,  since actual devices are
  annealed at \celcius{250} during fabrication.}
% This could be an effect of island growth in the thin CMG layer so
% that the CMG/RCS interface is partly missing giving a more ordered
% state for Cu, which fits well with the non-ferromagnetic state for
% thin CMG layers. While a thin RCS layer creates more disorder in the
% Mn
% component. %and also a bulk like component when part of a thin layer structure.
% We also notice that both Mn and Cu are sensitive to heat treatments
% above 200 C. The Mn is more sensitive to heat treatment and an
% effect can be observed already at 200 C.  The Rh is insensitive to
% the thickness of the magnetic layer and together with results from
% Monte Carlo simulations this suggests that the Rh is deficient at
% the interface.

\begin{acknowledgments}

  This work was supported by the Swedish Research Council (VR) and the
  Swedish Foundation for International Cooperation in Research and
  Higher Education (STINT). VR is also acknowledged for their financial support of SuperAdam. We acknowledge the Helmholtz-Zentrum
  Berlin - Electron storage ring BESSY II for provision of synchrotron
  radiation at beamline KMC-1. We also would like to thank M.Mertin for
  assistance. The research leading to these results has received
  funding from the European Community's Seventh Framework Programme
  (FP7/2007-2013) under grant agreement n.\ $^{\circ}226716$. Olof Karis gratefully acknowledge the support of the G{\"o}ran
  Gustavsson Foundation.
\end{acknowledgments}

\bibliographystyle{apsrev4-1} \bibliography{CMG_refs}

\begin{thebibliography}{34}%
\makeatletter
\providecommand \@ifxundefined [1]{%
 \@ifx{#1\undefined}
}%
\providecommand \@ifnum [1]{%
 \ifnum #1\expandafter \@firstoftwo
 \else \expandafter \@secondoftwo
 \fi
}%
\providecommand \@ifx [1]{%
 \ifx #1\expandafter \@firstoftwo
 \else \expandafter \@secondoftwo
 \fi
}%
\providecommand \natexlab [1]{#1}%
\providecommand \enquote  [1]{``#1''}%
\providecommand \bibnamefont  [1]{#1}%
\providecommand \bibfnamefont [1]{#1}%
\providecommand \citenamefont [1]{#1}%
\providecommand \href@noop [0]{\@secondoftwo}%
\providecommand \href [0]{\begingroup \@sanitize@url \@href}%
\providecommand \@href[1]{\@@startlink{#1}\@@href}%
\providecommand \@@href[1]{\endgroup#1\@@endlink}%
\providecommand \@sanitize@url [0]{\catcode `\\12\catcode `\$12\catcode
  `\&12\catcode `\#12\catcode `\^12\catcode `\_12\catcode `\%12\relax}%
\providecommand \@@startlink[1]{}%
\providecommand \@@endlink[0]{}%
\providecommand \url  [0]{\begingroup\@sanitize@url \@url }%
\providecommand \@url [1]{\endgroup\@href {#1}{\urlprefix }}%
\providecommand \urlprefix  [0]{URL }%
\providecommand \Eprint [0]{\href }%
\providecommand \doibase [0]{http://dx.doi.org/}%
\providecommand \selectlanguage [0]{\@gobble}%
\providecommand \bibinfo  [0]{\@secondoftwo}%
\providecommand \bibfield  [0]{\@secondoftwo}%
\providecommand \translation [1]{[#1]}%
\providecommand \BibitemOpen [0]{}%
\providecommand \bibitemStop [0]{}%
\providecommand \bibitemNoStop [0]{.\EOS\space}%
\providecommand \EOS [0]{\spacefactor3000\relax}%
\providecommand \BibitemShut  [1]{\csname bibitem#1\endcsname}%
\let\auto@bib@innerbib\@empty
%</preamble>
\bibitem [{\citenamefont {Baibich}\ \emph {et~al.}(1988)\citenamefont
  {Baibich}, \citenamefont {Broto}, \citenamefont {Fert}, \citenamefont
  {Vandau}, \citenamefont {Petroff}, \citenamefont {Eitenne}, \citenamefont
  {Creuzet}, \citenamefont {Friederich},\ and\ \citenamefont
  {Chazelas}}]{BAIBICH:1988ys}%
  \BibitemOpen
  \bibfield  {author} {\bibinfo {author} {\bibfnamefont {M.}~\bibnamefont
  {Baibich}}, \bibinfo {author} {\bibfnamefont {J.}~\bibnamefont {Broto}},
  \bibinfo {author} {\bibfnamefont {A.}~\bibnamefont {Fert}}, \bibinfo {author}
  {\bibfnamefont {F.}~\bibnamefont {Vandau}}, \bibinfo {author} {\bibfnamefont
  {F.}~\bibnamefont {Petroff}}, \bibinfo {author} {\bibfnamefont
  {P.}~\bibnamefont {Eitenne}}, \bibinfo {author} {\bibfnamefont
  {G.}~\bibnamefont {Creuzet}}, \bibinfo {author} {\bibfnamefont
  {A.}~\bibnamefont {Friederich}}, \ and\ \bibinfo {author} {\bibfnamefont
  {J.}~\bibnamefont {Chazelas}},\ }\href@noop {} {\bibfield  {journal}
  {\bibinfo  {journal} {Phys. Rev. Lett.}\ }\textbf {\bibinfo {volume} {61}},\
  \bibinfo {pages} {2472} (\bibinfo {year} {1988})}\BibitemShut {NoStop}%
\bibitem [{\citenamefont {Binasch}\ \emph {et~al.}(1989)\citenamefont
  {Binasch}, \citenamefont {Gr{\"u}nberg}, \citenamefont {Saurenbach},\ and\
  \citenamefont {Zinn}}]{BINASCH:1989vn}%
  \BibitemOpen
  \bibfield  {author} {\bibinfo {author} {\bibfnamefont {G.}~\bibnamefont
  {Binasch}}, \bibinfo {author} {\bibfnamefont {P.}~\bibnamefont
  {Gr{\"u}nberg}}, \bibinfo {author} {\bibfnamefont {F.}~\bibnamefont
  {Saurenbach}}, \ and\ \bibinfo {author} {\bibfnamefont {W.}~\bibnamefont
  {Zinn}},\ }\href@noop {} {\bibfield  {journal} {\bibinfo  {journal} {Phys.
  Rev. B}\ }\textbf {\bibinfo {volume} {39}},\ \bibinfo {pages} {4828}
  (\bibinfo {year} {1989})}\BibitemShut {NoStop}%
\bibitem [{\citenamefont {Takagishi}\ \emph {et~al.}(2010)\citenamefont
  {Takagishi}, \citenamefont {Yamada}, \citenamefont {Iwasaki}, \citenamefont
  {Fuke},\ and\ \citenamefont {Hashimoto}}]{Takagishi2010}%
  \BibitemOpen
  \bibfield  {author} {\bibinfo {author} {\bibfnamefont {M.}~\bibnamefont
  {Takagishi}}, \bibinfo {author} {\bibfnamefont {K.}~\bibnamefont {Yamada}},
  \bibinfo {author} {\bibfnamefont {H.}~\bibnamefont {Iwasaki}}, \bibinfo
  {author} {\bibfnamefont {H.~N.}\ \bibnamefont {Fuke}}, \ and\ \bibinfo
  {author} {\bibfnamefont {S.}~\bibnamefont {Hashimoto}},\ }\href@noop {}
  {\bibfield  {journal} {\bibinfo  {journal} {IEEE Transactions on Magnetics}\
  }\textbf {\bibinfo {volume} {46}},\ \bibinfo {pages} {2086} (\bibinfo {year}
  {2010})}\BibitemShut {NoStop}%
\bibitem [{\citenamefont {de~Groot}\ \emph {et~al.}(1983)\citenamefont
  {de~Groot}, \citenamefont {Mueller}, \citenamefont {Engen},\ and\
  \citenamefont {Buschow}}]{deGroot_PRL_1985}%
  \BibitemOpen
  \bibfield  {author} {\bibinfo {author} {\bibfnamefont {R.~A.}\ \bibnamefont
  {de~Groot}}, \bibinfo {author} {\bibfnamefont {F.~M.}\ \bibnamefont
  {Mueller}}, \bibinfo {author} {\bibfnamefont {P.~G.~v.}\ \bibnamefont
  {Engen}}, \ and\ \bibinfo {author} {\bibfnamefont {K.~H.~J.}\ \bibnamefont
  {Buschow}},\ }\href {\doibase 10.1103/PhysRevLett.50.2024} {\bibfield
  {journal} {\bibinfo  {journal} {Phys. Rev. Lett.}\ }\textbf {\bibinfo
  {volume} {50}},\ \bibinfo {pages} {2024} (\bibinfo {year}
  {1983})}\BibitemShut {NoStop}%
\bibitem [{\citenamefont {Katsnelson}\ \emph {et~al.}(2008)\citenamefont
  {Katsnelson}, \citenamefont {Irkhin}, \citenamefont {Chioncel}, \citenamefont
  {Lichtenstein},\ and\ \citenamefont {de~Groot}}]{Katsnelson_RevModPhys_2008}%
  \BibitemOpen
  \bibfield  {author} {\bibinfo {author} {\bibfnamefont {M.~I.}\ \bibnamefont
  {Katsnelson}}, \bibinfo {author} {\bibfnamefont {V.~Y.}\ \bibnamefont
  {Irkhin}}, \bibinfo {author} {\bibfnamefont {L.}~\bibnamefont {Chioncel}},
  \bibinfo {author} {\bibfnamefont {A.~I.}\ \bibnamefont {Lichtenstein}}, \
  and\ \bibinfo {author} {\bibfnamefont {R.~A.}\ \bibnamefont {de~Groot}},\
  }\href {\doibase 10.1103/RevModPhys.80.315} {\bibfield  {journal} {\bibinfo
  {journal} {Rev. Mod. Phys.}\ }\textbf {\bibinfo {volume} {80}},\ \bibinfo
  {pages} {315} (\bibinfo {year} {2008})}\BibitemShut {NoStop}%
\bibitem [{\citenamefont {Caballero}\ \emph {et~al.}(1998)\citenamefont
  {Caballero}, \citenamefont {Park}, \citenamefont {Childress}, \citenamefont
  {Bass}, \citenamefont {Chiang}, \citenamefont {Reilly}, \citenamefont
  {Pratt},\ and\ \citenamefont {Petroff}}]{Caballero19981801}%
  \BibitemOpen
  \bibfield  {author} {\bibinfo {author} {\bibfnamefont {J.}~\bibnamefont
  {Caballero}}, \bibinfo {author} {\bibfnamefont {Y.}~\bibnamefont {Park}},
  \bibinfo {author} {\bibfnamefont {J.}~\bibnamefont {Childress}}, \bibinfo
  {author} {\bibfnamefont {J.}~\bibnamefont {Bass}}, \bibinfo {author}
  {\bibfnamefont {W.}~\bibnamefont {Chiang}}, \bibinfo {author} {\bibfnamefont
  {A.}~\bibnamefont {Reilly}}, \bibinfo {author} {\bibfnamefont
  {W.}~\bibnamefont {Pratt}}, \ and\ \bibinfo {author} {\bibfnamefont
  {F.}~\bibnamefont {Petroff}},\ }\href@noop {} {\bibfield  {journal} {\bibinfo
   {journal} {J. Vac. Sci. \& Techn. A-Vacuum Surfaces and Films}\ }\textbf
  {\bibinfo {volume} {16}},\ \bibinfo {pages} {1801} (\bibinfo {year}
  {1998})}\BibitemShut {NoStop}%
\bibitem [{\citenamefont {Iwase}\ \emph {et~al.}(2009)\citenamefont {Iwase},
  \citenamefont {Sakuraba}, \citenamefont {Bosu}, \citenamefont {Saito},
  \citenamefont {Mitani},\ and\ \citenamefont {Takanashi}}]{APEX.2.063003}%
  \BibitemOpen
  \bibfield  {author} {\bibinfo {author} {\bibfnamefont {T.}~\bibnamefont
  {Iwase}}, \bibinfo {author} {\bibfnamefont {Y.}~\bibnamefont {Sakuraba}},
  \bibinfo {author} {\bibfnamefont {S.}~\bibnamefont {Bosu}}, \bibinfo {author}
  {\bibfnamefont {K.}~\bibnamefont {Saito}}, \bibinfo {author} {\bibfnamefont
  {S.}~\bibnamefont {Mitani}}, \ and\ \bibinfo {author} {\bibfnamefont
  {K.}~\bibnamefont {Takanashi}},\ }\href {\doibase 10.1143/APEX.2.063003}
  {\bibfield  {journal} {\bibinfo  {journal} {Appl. Phys. Expr.}\ }\textbf
  {\bibinfo {volume} {2}},\ \bibinfo {pages} {063003} (\bibinfo {year}
  {2009})}\BibitemShut {NoStop}%
\bibitem [{\citenamefont {Nakatani}\ \emph {et~al.}(2010)\citenamefont
  {Nakatani}, \citenamefont {Furubayashi}, \citenamefont {Kasai}, \citenamefont
  {Sukegawa}, \citenamefont {Takahashi}, \citenamefont {Mitani},\ and\
  \citenamefont {Hono}}]{nakatani:212501}%
  \BibitemOpen
  \bibfield  {author} {\bibinfo {author} {\bibfnamefont {T.~M.}\ \bibnamefont
  {Nakatani}}, \bibinfo {author} {\bibfnamefont {T.}~\bibnamefont
  {Furubayashi}}, \bibinfo {author} {\bibfnamefont {S.}~\bibnamefont {Kasai}},
  \bibinfo {author} {\bibfnamefont {H.}~\bibnamefont {Sukegawa}}, \bibinfo
  {author} {\bibfnamefont {Y.~K.}\ \bibnamefont {Takahashi}}, \bibinfo {author}
  {\bibfnamefont {S.}~\bibnamefont {Mitani}}, \ and\ \bibinfo {author}
  {\bibfnamefont {K.}~\bibnamefont {Hono}},\ }\href {\doibase
  10.1063/1.3432070} {\bibfield  {journal} {\bibinfo  {journal} {Appl. Phys.
  Lett.}\ }\textbf {\bibinfo {volume} {96}},\ \bibinfo {eid} {212501} (\bibinfo
  {year} {2010})}\BibitemShut {NoStop}%
\bibitem [{\citenamefont {Ambrose}\ and\ \citenamefont
  {Mryasov}(2005{\natexlab{a}})}]{Ambrose_Mryasov}%
  \BibitemOpen
  \bibfield  {author} {\bibinfo {author} {\bibfnamefont {T.}~\bibnamefont
  {Ambrose}}\ and\ \bibinfo {author} {\bibfnamefont {O.}~\bibnamefont
  {Mryasov}},\ }\enquote {\bibinfo {title} {Half-metallic alloys: Fundamentals
  and applications},}\ \ (\bibinfo  {publisher} {Springer Verlag, Berlin-New
  York},\ \bibinfo {year} {2005})\BibitemShut {NoStop}%
\bibitem [{\citenamefont {Ambrose}\ and\ \citenamefont
  {Mryasov}(2005{\natexlab{b}})}]{Ambrose_Mryasov_Patent}%
  \BibitemOpen
  \bibfield  {author} {\bibinfo {author} {\bibfnamefont {T.}~\bibnamefont
  {Ambrose}}\ and\ \bibinfo {author} {\bibfnamefont {O.}~\bibnamefont
  {Mryasov}},\ }\href@noop {} {\bibfield  {journal} {\bibinfo  {journal} {U.S.
  Patent No. 6,876,522}\ } (\bibinfo {year} {5 April
  2005}{\natexlab{b}})}\BibitemShut {NoStop}%
\bibitem [{\citenamefont {Nikolaev}\ \emph {et~al.}(2009)\citenamefont
  {Nikolaev}, \citenamefont {Kolbo}, \citenamefont {Pokhil}, \citenamefont
  {Peng}, \citenamefont {Chen}, \citenamefont {Ambrose},\ and\ \citenamefont
  {Mryasov}}]{Mryasov_APL_2009}%
  \BibitemOpen
  \bibfield  {author} {\bibinfo {author} {\bibfnamefont {K.}~\bibnamefont
  {Nikolaev}}, \bibinfo {author} {\bibfnamefont {P.}~\bibnamefont {Kolbo}},
  \bibinfo {author} {\bibfnamefont {T.}~\bibnamefont {Pokhil}}, \bibinfo
  {author} {\bibfnamefont {X.}~\bibnamefont {Peng}}, \bibinfo {author}
  {\bibfnamefont {Y.}~\bibnamefont {Chen}}, \bibinfo {author} {\bibfnamefont
  {T.}~\bibnamefont {Ambrose}}, \ and\ \bibinfo {author} {\bibfnamefont
  {O.}~\bibnamefont {Mryasov}},\ }\href {http://dx.doi.org/10.1063/1.3126962}
  {\bibfield  {journal} {\bibinfo  {journal} {Appl. Phys. Lett.}\ }\textbf
  {\bibinfo {volume} {94}},\ \bibinfo {pages} {222501} (\bibinfo {year}
  {2009})}\BibitemShut {NoStop}%
\bibitem [{\citenamefont {Sato}\ \emph {et~al.}(2011)\citenamefont {Sato},
  \citenamefont {Oogane}, \citenamefont {Naganuma},\ and\ \citenamefont
  {Ando}}]{APEX.4.113005}%
  \BibitemOpen
  \bibfield  {author} {\bibinfo {author} {\bibfnamefont {J.}~\bibnamefont
  {Sato}}, \bibinfo {author} {\bibfnamefont {M.}~\bibnamefont {Oogane}},
  \bibinfo {author} {\bibfnamefont {H.}~\bibnamefont {Naganuma}}, \ and\
  \bibinfo {author} {\bibfnamefont {Y.}~\bibnamefont {Ando}},\ }\href {\doibase
  10.1143/APEX.4.113005} {\bibfield  {journal} {\bibinfo  {journal} {Appl.
  Phys. Expr.}\ }\textbf {\bibinfo {volume} {4}},\ \bibinfo {pages} {113005}
  (\bibinfo {year} {2011})}\BibitemShut {NoStop}%
\bibitem [{\citenamefont {Carey}(2004)}]{Carey_APL_2004}%
  \BibitemOpen
  \bibfield  {author} {\bibinfo {author} {\bibfnamefont {M.}~\bibnamefont
  {Carey}},\ }\href {http://dx.doi.org/10.1063/1.1818735} {\bibfield  {journal}
  {\bibinfo  {journal} {et al, Appl. Phys. Lett.}\ }\textbf {\bibinfo {volume}
  {85}},\ \bibinfo {pages} {4442 } (\bibinfo {year} {2004})}\BibitemShut
  {NoStop}%
\bibitem [{\citenamefont {Ozdogan}\ and\ \citenamefont
  {Galanakis}(2011)}]{Ozdogan:2011qy}%
  \BibitemOpen
  \bibfield  {author} {\bibinfo {author} {\bibfnamefont {K.}~\bibnamefont
  {Ozdogan}}\ and\ \bibinfo {author} {\bibfnamefont {I.}~\bibnamefont
  {Galanakis}},\ }\href {\doibase 10.1063/1.3642990} {\bibfield  {journal}
  {\bibinfo  {journal} {J. Appl. Phys.}\ }\textbf {\bibinfo {volume} {110}},\
  \bibinfo {pages} {076101} (\bibinfo {year} {2011})}\BibitemShut {NoStop}%
\bibitem [{\citenamefont {Galanakis}\ \emph {et~al.}(2009)\citenamefont
  {Galanakis}, \citenamefont {Oezdogan},\ and\ \citenamefont
  {Sasioglu}}]{Galanakis:2009rt}%
  \BibitemOpen
  \bibfield  {author} {\bibinfo {author} {\bibfnamefont {I.}~\bibnamefont
  {Galanakis}}, \bibinfo {author} {\bibfnamefont {K.}~\bibnamefont {Oezdogan}},
  \ and\ \bibinfo {author} {\bibfnamefont {E.}~\bibnamefont {Sasioglu}},\ }in\
  \href@noop {} {\emph {\bibinfo {booktitle} {Advances In Nanoscale
  Magnetism}}},\ \bibinfo {series} {Springer Proceedings in Physics}, Vol.\
  \bibinfo {volume} {122}\ (\bibinfo  {publisher} {Springer-Verlag},\ \bibinfo
  {address} {Berlin, Germany},\ \bibinfo {year} {2009})\ p.\ \bibinfo {pages}
  {328}\BibitemShut {NoStop}%
\bibitem [{\citenamefont {Ravel}\ \emph {et~al.}(2002)\citenamefont {Ravel},
  \citenamefont {Cross}, \citenamefont {Raphael}, \citenamefont {Harris},
  \citenamefont {Ramesh},\ and\ \citenamefont {Saraf}}]{Ravel20022812}%
  \BibitemOpen
  \bibfield  {author} {\bibinfo {author} {\bibfnamefont {B.}~\bibnamefont
  {Ravel}}, \bibinfo {author} {\bibfnamefont {J.}~\bibnamefont {Cross}},
  \bibinfo {author} {\bibfnamefont {M.}~\bibnamefont {Raphael}}, \bibinfo
  {author} {\bibfnamefont {V.}~\bibnamefont {Harris}}, \bibinfo {author}
  {\bibfnamefont {R.}~\bibnamefont {Ramesh}}, \ and\ \bibinfo {author}
  {\bibfnamefont {V.}~\bibnamefont {Saraf}},\ }\href
  {http://www.scopus.com/inward/record.url?eid=2-s2.0-79956041724&partnerID=40&md5=63025c6810fcbffd4fd90f27bdf3e6d5}
  {\bibfield  {journal} {\bibinfo  {journal} {Appl. Phys. Lett.}\ }\textbf
  {\bibinfo {volume} {81}},\ \bibinfo {pages} {2812} (\bibinfo {year}
  {2002})},\ \bibinfo {note} {cited By (since 1996) 47}\BibitemShut {NoStop}%
\bibitem [{\citenamefont {Lezaic}\ \emph {et~al.}(2011)\citenamefont {Lezaic},
  \citenamefont {Mavropoulos}, \citenamefont {Bluegel},\ and\ \citenamefont
  {Ebert}}]{Lezaic:2011fj}%
  \BibitemOpen
  \bibfield  {author} {\bibinfo {author} {\bibfnamefont {M.}~\bibnamefont
  {Lezaic}}, \bibinfo {author} {\bibfnamefont {P.}~\bibnamefont {Mavropoulos}},
  \bibinfo {author} {\bibfnamefont {S.}~\bibnamefont {Bluegel}}, \ and\
  \bibinfo {author} {\bibfnamefont {H.}~\bibnamefont {Ebert}},\ }\href
  {\doibase DOI 10.1103/PhysRevB.83.094434} {\bibfield  {journal} {\bibinfo
  {journal} {Phys. Rev. B}\ }\textbf {\bibinfo {volume} {83}},\ \bibinfo
  {pages} {094434} (\bibinfo {year} {2011})}\BibitemShut {NoStop}%
\bibitem [{\citenamefont {Carra}\ \emph {et~al.}(1993)\citenamefont {Carra},
  \citenamefont {Thole}, \citenamefont {Altarelli},\ and\ \citenamefont
  {Wang}}]{PhysRevLett.70.694}%
  \BibitemOpen
  \bibfield  {author} {\bibinfo {author} {\bibfnamefont {P.}~\bibnamefont
  {Carra}}, \bibinfo {author} {\bibfnamefont {B.~T.}\ \bibnamefont {Thole}},
  \bibinfo {author} {\bibfnamefont {M.}~\bibnamefont {Altarelli}}, \ and\
  \bibinfo {author} {\bibfnamefont {X.}~\bibnamefont {Wang}},\ }\href {\doibase
  10.1103/PhysRevLett.70.694} {\bibfield  {journal} {\bibinfo  {journal} {Phys.
  Rev. Lett.}\ }\textbf {\bibinfo {volume} {70}},\ \bibinfo {pages} {694}
  (\bibinfo {year} {1993})}\BibitemShut {NoStop}%
\bibitem [{\citenamefont {Thole}\ \emph {et~al.}(1992)\citenamefont {Thole},
  \citenamefont {Carra}, \citenamefont {Sette},\ and\ \citenamefont {van~der
  Laan}}]{PhysRevLett.68.1943}%
  \BibitemOpen
  \bibfield  {author} {\bibinfo {author} {\bibfnamefont {B.~T.}\ \bibnamefont
  {Thole}}, \bibinfo {author} {\bibfnamefont {P.}~\bibnamefont {Carra}},
  \bibinfo {author} {\bibfnamefont {F.}~\bibnamefont {Sette}}, \ and\ \bibinfo
  {author} {\bibfnamefont {G.}~\bibnamefont {van~der Laan}},\ }\href {\doibase
  10.1103/PhysRevLett.68.1943} {\bibfield  {journal} {\bibinfo  {journal}
  {Phys. Rev. Lett.}\ }\textbf {\bibinfo {volume} {68}},\ \bibinfo {pages}
  {1943} (\bibinfo {year} {1992})}\BibitemShut {NoStop}%
\bibitem [{\citenamefont {Holmstr{\"o}m}\ \emph {et~al.}(2006)\citenamefont
  {Holmstr{\"o}m}, \citenamefont {Olovsson}, \citenamefont {Abrikosov},
  \citenamefont {Niklasson}, \citenamefont {Johansson}, \citenamefont {Gorgoi},
  \citenamefont {Karis}, \citenamefont {Svensson}, \citenamefont
  {Sch{\"a}fers}, \citenamefont {Braun}, \citenamefont {{\"O}hrwall},
  \citenamefont {Andersson}, \citenamefont {Marcellini},\ and\ \citenamefont
  {Eberhardt}}]{holmstrom2006}%
  \BibitemOpen
  \bibfield  {author} {\bibinfo {author} {\bibfnamefont {E.}~\bibnamefont
  {Holmstr{\"o}m}}, \bibinfo {author} {\bibfnamefont {W.}~\bibnamefont
  {Olovsson}}, \bibinfo {author} {\bibfnamefont {I.}~\bibnamefont {Abrikosov}},
  \bibinfo {author} {\bibfnamefont {A.}~\bibnamefont {Niklasson}}, \bibinfo
  {author} {\bibfnamefont {B.}~\bibnamefont {Johansson}}, \bibinfo {author}
  {\bibfnamefont {M.}~\bibnamefont {Gorgoi}}, \bibinfo {author} {\bibfnamefont
  {O.}~\bibnamefont {Karis}}, \bibinfo {author} {\bibfnamefont
  {S.}~\bibnamefont {Svensson}}, \bibinfo {author} {\bibfnamefont
  {F.}~\bibnamefont {Sch{\"a}fers}}, \bibinfo {author} {\bibfnamefont
  {W.}~\bibnamefont {Braun}}, \bibinfo {author} {\bibfnamefont
  {G.}~\bibnamefont {{\"O}hrwall}}, \bibinfo {author} {\bibfnamefont
  {G.}~\bibnamefont {Andersson}}, \bibinfo {author} {\bibfnamefont
  {M.}~\bibnamefont {Marcellini}}, \ and\ \bibinfo {author} {\bibfnamefont
  {W.}~\bibnamefont {Eberhardt}},\ }\href
  {http://dx.doi.org/10.1103/PhysRevLett.97.266106} {\bibfield  {journal}
  {\bibinfo  {journal} {Phys. Rev. Lett.}\ }\textbf {\bibinfo {volume} {97}},\
  \bibinfo {pages} {266106} (\bibinfo {year} {2006})}\BibitemShut {NoStop}%
\bibitem [{\citenamefont {Granroth}\ \emph {et~al.}(2009)\citenamefont
  {Granroth}, \citenamefont {Knut}, \citenamefont {Marcellini}, \citenamefont
  {Andersson}, \citenamefont {Svensson}, \citenamefont {Karis}, \citenamefont
  {Gorgoi}, \citenamefont {Sch{\"a}fers}, \citenamefont {Braun}, \citenamefont
  {Eberhardt}, \citenamefont {Olovsson}, \citenamefont {Holmstr{\"o}m},\ and\
  \citenamefont {M{\aa}rtensson}}]{granroth:094104}%
  \BibitemOpen
  \bibfield  {author} {\bibinfo {author} {\bibfnamefont {S.}~\bibnamefont
  {Granroth}}, \bibinfo {author} {\bibfnamefont {R.}~\bibnamefont {Knut}},
  \bibinfo {author} {\bibfnamefont {M.}~\bibnamefont {Marcellini}}, \bibinfo
  {author} {\bibfnamefont {G.}~\bibnamefont {Andersson}}, \bibinfo {author}
  {\bibfnamefont {S.}~\bibnamefont {Svensson}}, \bibinfo {author}
  {\bibfnamefont {O.}~\bibnamefont {Karis}}, \bibinfo {author} {\bibfnamefont
  {M.}~\bibnamefont {Gorgoi}}, \bibinfo {author} {\bibfnamefont
  {F.}~\bibnamefont {Sch{\"a}fers}}, \bibinfo {author} {\bibfnamefont
  {W.}~\bibnamefont {Braun}}, \bibinfo {author} {\bibfnamefont
  {W.}~\bibnamefont {Eberhardt}}, \bibinfo {author} {\bibfnamefont
  {W.}~\bibnamefont {Olovsson}}, \bibinfo {author} {\bibfnamefont
  {E.}~\bibnamefont {Holmstr{\"o}m}}, \ and\ \bibinfo {author} {\bibfnamefont
  {N.}~\bibnamefont {M{\aa}rtensson}},\ }\href {\doibase
  10.1103/PhysRevB.80.094104} {\bibfield  {journal} {\bibinfo  {journal} {Phys.
  Rev. B}\ }\textbf {\bibinfo {volume} {80}},\ \bibinfo {eid} {094104}
  (\bibinfo {year} {2009})}\BibitemShut {NoStop}%
\bibitem [{\citenamefont {Gorgoi}\ \emph {et~al.}(2009)\citenamefont {Gorgoi},
  \citenamefont {Svensson}, \citenamefont {Sch{\"a}fers}, \citenamefont
  {{\"O}hrwall}, \citenamefont {Mertin}, \citenamefont {Bressler},
  \citenamefont {Karis}, \citenamefont {Siegbahn}, \citenamefont {Sandell},
  \citenamefont {Rensmo}, \citenamefont {Doherty}, \citenamefont {Jung},
  \citenamefont {Braun},\ and\ \citenamefont {Eberhardt}}]{unp-go.sv.ea:08}%
  \BibitemOpen
  \bibfield  {author} {\bibinfo {author} {\bibfnamefont {M.}~\bibnamefont
  {Gorgoi}}, \bibinfo {author} {\bibfnamefont {S.}~\bibnamefont {Svensson}},
  \bibinfo {author} {\bibfnamefont {F.}~\bibnamefont {Sch{\"a}fers}}, \bibinfo
  {author} {\bibfnamefont {G.}~\bibnamefont {{\"O}hrwall}}, \bibinfo {author}
  {\bibfnamefont {M.}~\bibnamefont {Mertin}}, \bibinfo {author} {\bibfnamefont
  {P.}~\bibnamefont {Bressler}}, \bibinfo {author} {\bibfnamefont
  {O.}~\bibnamefont {Karis}}, \bibinfo {author} {\bibfnamefont
  {H.}~\bibnamefont {Siegbahn}}, \bibinfo {author} {\bibfnamefont
  {A.}~\bibnamefont {Sandell}}, \bibinfo {author} {\bibfnamefont
  {H.}~\bibnamefont {Rensmo}}, \bibinfo {author} {\bibfnamefont
  {W.}~\bibnamefont {Doherty}}, \bibinfo {author} {\bibfnamefont
  {C.}~\bibnamefont {Jung}}, \bibinfo {author} {\bibfnamefont {W.}~\bibnamefont
  {Braun}}, \ and\ \bibinfo {author} {\bibfnamefont {W.}~\bibnamefont
  {Eberhardt}},\ }\href {\doibase DOI: 10.1016/j.nima.2008.12.244} {\bibfield
  {journal} {\bibinfo  {journal} {Nuclear Instruments and Methods in Physics
  Research Section A: Accelerators, Spectrometers, Detectors and Associated
  Equipment}\ }\textbf {\bibinfo {volume} {601}},\ \bibinfo {pages} {48 }
  (\bibinfo {year} {2009})},\ \bibinfo {note} {special issue in honour of Prof.
  Kai Siegbahn}\BibitemShut {NoStop}%
\bibitem [{\citenamefont {Knut}\ \emph {et~al.}()\citenamefont {Knut} \emph
  {et~al.}}]{Ronny_Heusler_Unpub}%
  \BibitemOpen
  \bibfield  {author} {\bibinfo {author} {\bibfnamefont {R.}~\bibnamefont
  {Knut}} \emph {et~al.},\ }\href@noop {} {}\bibinfo {note} {To be
  published}\BibitemShut {NoStop}%
\bibitem [{\citenamefont {Johansson}\ and\ \citenamefont
  {M{\aa}rtensson}(1980)}]{Johansson:1980fk}%
  \BibitemOpen
  \bibfield  {author} {\bibinfo {author} {\bibfnamefont {B.}~\bibnamefont
  {Johansson}}\ and\ \bibinfo {author} {\bibfnamefont {N.}~\bibnamefont
  {M{\aa}rtensson}},\ }\href {\doibase 10.1103/PhysRevB.21.4427} {\bibfield
  {journal} {\bibinfo  {journal} {Phys. Rev. B}\ }\textbf {\bibinfo {volume}
  {21}},\ \bibinfo {pages} {4427} (\bibinfo {year} {1980})}\BibitemShut
  {NoStop}%
\bibitem [{\citenamefont {Doniach}\ and\ \citenamefont
  {Sunjic}(1970)}]{Doniach:1970fk}%
  \BibitemOpen
  \bibfield  {author} {\bibinfo {author} {\bibfnamefont {S.}~\bibnamefont
  {Doniach}}\ and\ \bibinfo {author} {\bibfnamefont {M.}~\bibnamefont
  {Sunjic}},\ }\href {http://stacks.iop.org/0022-3719/3/i=2/a=010} {\bibfield
  {journal} {\bibinfo  {journal} {J. Phys. C: Solid State Physics}\ }\textbf
  {\bibinfo {volume} {3}},\ \bibinfo {pages} {285} (\bibinfo {year}
  {1970})}\BibitemShut {NoStop}%
\bibitem [{\citenamefont {V{\'e}gh}(2006)}]{Vegh:2006fk}%
  \BibitemOpen
  \bibfield  {author} {\bibinfo {author} {\bibfnamefont {J.}~\bibnamefont
  {V{\'e}gh}},\ }\href {\doibase 10.1016/j.elspec.2005.12.002} {\bibfield
  {journal} {\bibinfo  {journal} {Journal of Electron Spectroscopy and Related
  Phenomena}\ }\textbf {\bibinfo {volume} {151}},\ \bibinfo {pages} {159}
  (\bibinfo {year} {2006})},\ \bibinfo {note} {and references
  therein.}\BibitemShut {Stop}%
\bibitem [{\citenamefont {Varaprasad}\ \emph {et~al.}(2010)\citenamefont
  {Varaprasad}, \citenamefont {Rajanikanth}, \citenamefont {Takahashi},\ and\
  \citenamefont {Hono}}]{Varaprasad_APEx_2010}%
  \BibitemOpen
  \bibfield  {author} {\bibinfo {author} {\bibfnamefont {B.}~\bibnamefont
  {Varaprasad}}, \bibinfo {author} {\bibfnamefont {A.}~\bibnamefont
  {Rajanikanth}}, \bibinfo {author} {\bibfnamefont {Y.}~\bibnamefont
  {Takahashi}}, \ and\ \bibinfo {author} {\bibfnamefont {K.}~\bibnamefont
  {Hono}},\ }\href {http://dx.doi.org/10.1143/APEX.3.023002} {\bibfield
  {journal} {\bibinfo  {journal} {Appl. Phys. Exp.}\ }\textbf {\bibinfo
  {volume} {3}},\ \bibinfo {pages} {023002} (\bibinfo {year}
  {2010})}\BibitemShut {NoStop}%
\bibitem [{\citenamefont {Antonov}\ \emph {et~al.}(2006)\citenamefont
  {Antonov}, \citenamefont {Jepsen}, \citenamefont {Yaresko},\ and\
  \citenamefont {Shpak}}]{Antonov_JAP_2006}%
  \BibitemOpen
  \bibfield  {author} {\bibinfo {author} {\bibfnamefont {V.~N.}\ \bibnamefont
  {Antonov}}, \bibinfo {author} {\bibfnamefont {O.}~\bibnamefont {Jepsen}},
  \bibinfo {author} {\bibfnamefont {A.~N.}\ \bibnamefont {Yaresko}}, \ and\
  \bibinfo {author} {\bibfnamefont {A.~P.}\ \bibnamefont {Shpak}},\ }\href@noop
  {} {\bibfield  {journal} {\bibinfo  {journal} {J. Appl. Phys.}\ }\textbf
  {\bibinfo {volume} {{100}}} (\bibinfo {year} {2006})}\BibitemShut {NoStop}%
\bibitem [{\citenamefont {Grabis}\ \emph {et~al.}(2005)\citenamefont {Grabis},
  \citenamefont {Bergmann}, \citenamefont {Nefedov}, \citenamefont
  {Westerholt},\ and\ \citenamefont {Zabel}}]{PhysRevB.72.024437}%
  \BibitemOpen
  \bibfield  {author} {\bibinfo {author} {\bibfnamefont {J.}~\bibnamefont
  {Grabis}}, \bibinfo {author} {\bibfnamefont {A.}~\bibnamefont {Bergmann}},
  \bibinfo {author} {\bibfnamefont {A.}~\bibnamefont {Nefedov}}, \bibinfo
  {author} {\bibfnamefont {K.}~\bibnamefont {Westerholt}}, \ and\ \bibinfo
  {author} {\bibfnamefont {H.}~\bibnamefont {Zabel}},\ }\href {\doibase
  10.1103/PhysRevB.72.024437} {\bibfield  {journal} {\bibinfo  {journal} {Phys.
  Rev. B}\ }\textbf {\bibinfo {volume} {72}},\ \bibinfo {pages} {024437}
  (\bibinfo {year} {2005})}\BibitemShut {NoStop}%
\bibitem [{\citenamefont {Picozzi}\ \emph {et~al.}(2004)\citenamefont
  {Picozzi}, \citenamefont {Continenza},\ and\ \citenamefont
  {Freeman}}]{PhysRevB.69.094423}%
  \BibitemOpen
  \bibfield  {author} {\bibinfo {author} {\bibfnamefont {S.}~\bibnamefont
  {Picozzi}}, \bibinfo {author} {\bibfnamefont {A.}~\bibnamefont {Continenza}},
  \ and\ \bibinfo {author} {\bibfnamefont {A.~J.}\ \bibnamefont {Freeman}},\
  }\href {\doibase 10.1103/PhysRevB.69.094423} {\bibfield  {journal} {\bibinfo
  {journal} {Phys. Rev. B}\ }\textbf {\bibinfo {volume} {69}},\ \bibinfo
  {pages} {094423} (\bibinfo {year} {2004})}\BibitemShut {NoStop}%
\bibitem [{\citenamefont {Bj{\"o}rck}\ and\ \citenamefont
  {Andersson}(2007)}]{Bjorck20071174}%
  \BibitemOpen
  \bibfield  {author} {\bibinfo {author} {\bibfnamefont {M.}~\bibnamefont
  {Bj{\"o}rck}}\ and\ \bibinfo {author} {\bibfnamefont {G.}~\bibnamefont
  {Andersson}},\ }\href {\doibase 10.1107/S0021889807045086} {\bibfield
  {journal} {\bibinfo  {journal} {J. Appl. Crystallogr.}\ }\textbf {\bibinfo
  {volume} {40}},\ \bibinfo {pages} {1174} (\bibinfo {year}
  {2007})}\BibitemShut {NoStop}%
\bibitem [{\citenamefont {Asakura}\ \emph {et~al.}(2010)\citenamefont
  {Asakura}, \citenamefont {Koide}, \citenamefont {Yamamoto}, \citenamefont
  {Tsuchiya}, \citenamefont {Shioya}, \citenamefont {Amemiya}, \citenamefont
  {Singh}, \citenamefont {Kataoka}, \citenamefont {Yamazaki}, \citenamefont
  {Sakamoto}, \citenamefont {Fujimori}, \citenamefont {Taira},\ and\
  \citenamefont {Yamamoto}}]{PhysRevB.82.184419}%
  \BibitemOpen
  \bibfield  {author} {\bibinfo {author} {\bibfnamefont {D.}~\bibnamefont
  {Asakura}}, \bibinfo {author} {\bibfnamefont {T.}~\bibnamefont {Koide}},
  \bibinfo {author} {\bibfnamefont {S.}~\bibnamefont {Yamamoto}}, \bibinfo
  {author} {\bibfnamefont {K.}~\bibnamefont {Tsuchiya}}, \bibinfo {author}
  {\bibfnamefont {T.}~\bibnamefont {Shioya}}, \bibinfo {author} {\bibfnamefont
  {K.}~\bibnamefont {Amemiya}}, \bibinfo {author} {\bibfnamefont {V.~R.}\
  \bibnamefont {Singh}}, \bibinfo {author} {\bibfnamefont {T.}~\bibnamefont
  {Kataoka}}, \bibinfo {author} {\bibfnamefont {Y.}~\bibnamefont {Yamazaki}},
  \bibinfo {author} {\bibfnamefont {Y.}~\bibnamefont {Sakamoto}}, \bibinfo
  {author} {\bibfnamefont {A.}~\bibnamefont {Fujimori}}, \bibinfo {author}
  {\bibfnamefont {T.}~\bibnamefont {Taira}}, \ and\ \bibinfo {author}
  {\bibfnamefont {M.}~\bibnamefont {Yamamoto}},\ }\href {\doibase
  10.1103/PhysRevB.82.184419} {\bibfield  {journal} {\bibinfo  {journal} {Phys.
  Rev. B}\ }\textbf {\bibinfo {volume} {82}},\ \bibinfo {pages} {184419}
  (\bibinfo {year} {2010})}\BibitemShut {NoStop}%
\bibitem [{\citenamefont {Saito}\ \emph {et~al.}(2010)\citenamefont {Saito},
  \citenamefont {Katayama}, \citenamefont {Ishikawa}, \citenamefont {Yamamoto},
  \citenamefont {Asakura}, \citenamefont {Koide}, \citenamefont {Miura},\ and\
  \citenamefont {Shirai}}]{PhysRevB.81.144417}%
  \BibitemOpen
  \bibfield  {author} {\bibinfo {author} {\bibfnamefont {T.}~\bibnamefont
  {Saito}}, \bibinfo {author} {\bibfnamefont {T.}~\bibnamefont {Katayama}},
  \bibinfo {author} {\bibfnamefont {T.}~\bibnamefont {Ishikawa}}, \bibinfo
  {author} {\bibfnamefont {M.}~\bibnamefont {Yamamoto}}, \bibinfo {author}
  {\bibfnamefont {D.}~\bibnamefont {Asakura}}, \bibinfo {author} {\bibfnamefont
  {T.}~\bibnamefont {Koide}}, \bibinfo {author} {\bibfnamefont
  {Y.}~\bibnamefont {Miura}}, \ and\ \bibinfo {author} {\bibfnamefont
  {M.}~\bibnamefont {Shirai}},\ }\href {\doibase 10.1103/PhysRevB.81.144417}
  {\bibfield  {journal} {\bibinfo  {journal} {Phys. Rev. B}\ }\textbf {\bibinfo
  {volume} {81}},\ \bibinfo {pages} {144417} (\bibinfo {year}
  {2010})}\BibitemShut {NoStop}%
\bibitem [{\citenamefont {Teramura}\ \emph {et~al.}(1996)\citenamefont
  {Teramura}, \citenamefont {Tanaka},\ and\ \citenamefont {Jo}}]{JPSJ.65.1053}%
  \BibitemOpen
  \bibfield  {author} {\bibinfo {author} {\bibfnamefont {Y.}~\bibnamefont
  {Teramura}}, \bibinfo {author} {\bibfnamefont {A.}~\bibnamefont {Tanaka}}, \
  and\ \bibinfo {author} {\bibfnamefont {T.}~\bibnamefont {Jo}},\ }\href
  {\doibase 10.1143/JPSJ.65.1053} {\bibfield  {journal} {\bibinfo  {journal}
  {J. Phys. Soc. Jpn.}\ }\textbf {\bibinfo {volume} {65}},\ \bibinfo {pages}
  {1053} (\bibinfo {year} {1996})}\BibitemShut {NoStop}%
\end{thebibliography}%

\end{document}